\documentclass[twocolumn,showpacs,superscriptaddress,twoside,pra]{revtex4}
\usepackage{amsmath,amssymb,url}
\usepackage{graphicx,epstopdf,bm}
\usepackage{braket}
\usepackage{epsfig}
\usepackage{hyperref}
\usepackage{amsmath}
\usepackage{amssymb}
\usepackage{makeidx}
\usepackage{multirow}
\usepackage{xcolor}
\usepackage{color}
\usepackage[caption=false]{subfig}

\begin{document}
\title{ Slowing the probe field in the second window of 
	\\ double-double electromagnetically induced transparency}
\author{Hessa M. M. Alotaibi}
\email{hmalotai@ucalgary.ca}
\affiliation{Institute for Quantum Science and Technology,
	University of Calgary, Alberta, Canada T2N 1N4}
\affiliation{Public Authority for Applied Education and Training,
	P.O. Box 23167, Safat 13092, Kuwait }
\author{Barry C. Sanders}
\affiliation{Institute for Quantum Science and Technology, University of Calgary, Alberta, Canada T2N 1N4}
\affiliation{%
	Program in Quantum Information Science,
	Canadian Institute for Advanced Research, Toronto, Ontario M5G 1Z8, Canada
	}
\begin{abstract}
For Doppler-broadened media
operating under double-double electromagnetically induced transparency (EIT) conditions,
we devise a scheme to control and reduce the probe-field group velocity 
at the center of the second transparency window.
We derive numerical and approximate analytical solutions for the width of EIT windows and 
for the group velocities of
the probe field at the two distinct transparency windows,
and we show that the group velocities of the probe field 
can be lowered by judiciously choosing the physical parameters of the system.
Our modeling enables us to identify three signal-field strength regimes
(with a signal-field strength always higher than the probe-field strength),
quantified by the Rabi frequency,
for slowing the probe field.
These three regimes correspond to a weak signal field,
with the probe-field group velocity and transparency window width both smaller for the second window compared to the first window,
a medium-strength signal field,
with a probe-field group velocity smaller in the second window than in the first window
but with larger transparency-window width for the second window,
and the strong signal field,
with both group velocity and transparency window width larger for the second window.
Our scheme exploits the fact that the second transparency window is sensitive to a temperature-controlled signal-field nonlinearity, whereas the first transparency window is insensitive to this nonlinearity.
\end{abstract} 

\date{\today}
\pacs{42.50.Gy, 42.50.Ex}
\maketitle
\section{Introduction}
Electromagnetically induced transparency (EIT) is a phenomenon whereby a medium
can be switched between states of transparency and opacity through the controllable
application of a weak probe field~\cite{Harris1997},
and applications of EIT include slow light~\cite{Hau1999, Harris99,Kash1999}, optical switching~\cite{Harris1998}, and optical quantum memory~\cite{Lvovsky2009}.
Double EIT extends the notion of EIT from controllable creation of transparency at a given frequency 
to creation of two transparency windows at two different frequencies~\cite{Rebic2004}.

Double-double EIT (DDEIT) extends EIT even further to the case that each of a signal and a probe field
experience two transparency windows with the transparency window for the signal controlled by a coupling field and a probe field and the transparency window for the probe field controlled by the coupling field and the signal field~\cite{Hessa2013}.
DDEIT introduces the possibility of controlling propagation of and interaction between two bichromatic fields.

One advantage of controlling light is slowing it down.
Slow light is especially important for optical communication
and for quantum information processing.
For optical communication, slow light enhances light-matter interaction times and thereby
leads to an increase in nonlinear interactions~\cite{Harris99, Kash1999,Schmidt1996},.
In quantum computing, slow light enables storage of the quantum state of light for a sufficiently long time to enable quantum memory~\cite{Lvovsky2009}.

Our main objective is to propose a mechanism to slow the probe field in the second transparency window
of Doppler-broadened DDEIT.
For this slowing to be achieved,
we need to balance two competing requirements.
One requirement for slowing the probe pulse in 
Doppler-broadened EIT is to reduce the driving field intensity
because the group velocity is proportional the driving-field intensity.
On the other hand,
the driving-field intensity must be sufficiently large to circumvent
inhomogeneous broadening~\cite{Taichenachev2000, Yuri2002, Javan2002, Ye2002, Figueroa2006}.

We derive an analytical expression to enable us to find a parameter regime where these competing requirements can simultaneously be satisfied.
Our analytical technique is based on approximating the Maxwell-Boltzmann velocity
distribution for atoms by Lorentzian distributions over the narrow but relevant domain of small atomic velocities~\cite{Javan2002}.
We find that the nonlinear interaction between the probe and signal field maintain the width of the second window constant for high Doppler width. This result permits us to lower the intensity of the signal field further without losing the EIT  transparency window and get lower probe field group velocity at the second window compared to the first window.
We apply our scheme to the case of ${}^{87}$Rb
under realistic experimental conditions
and show that group-velocity reduction of the probe field is feasible.

We present our work in the following order.
In Sec.~\ref{sec:stationaryatom}, 
we introduce the optical density matrix element describing the optical properties of 
the transition $\ket{1}\rightarrow\ket{4}$ and the resultant atomic susceptibility,
and we use these results to calculate the widths
of the Lorentzian-shaped transparency windows and the 
corresponding group velocities for the probe field.

Doppler broadening due to temperature is incorporated into the expression for susceptibility in Sec.~\ref{sec:Doppler}. We solve this susceptibility numerically and analytically.
Our analytical solution is based on ignoring quadratic dependence of the probe-field Rabi
frequency and employing a Lorentzian approximation for a narrow 
band around the Gaussian Maxwell-Boltzmann distribution. This
approximate expression enables an intuition about how to control group velocities' reduction at the second window.
\begin{figure}
 \includegraphics [width=.7\columnwidth]{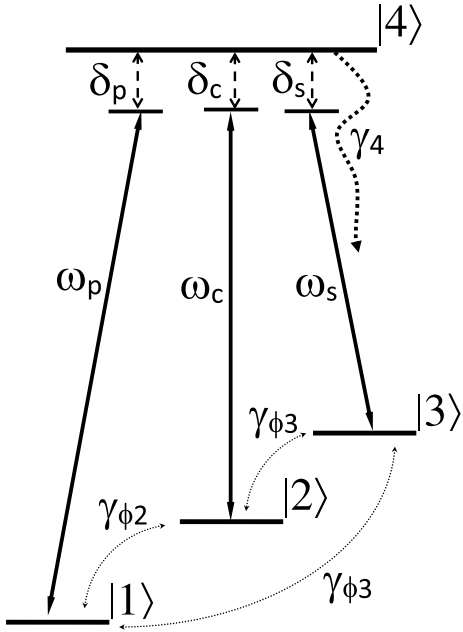}
 \caption{Four-level tripod electronic structure with high-energy state~$|4\rangle$
		and lower-energy levels~$\ket{3}$,~$\ket{2}$, and~$\ket{1}$ in order of 
		decreasing energy. Transitions are driven by probe~(p), coupling~(c) and 
	signal~(s) fields with frequencies $\omega_\text{x}$ and detunings $\delta_\text{x}$
		with x$\in\{\text{p,c,s}\}$. 
		Dephasing rates are $\gamma_{\phi i}$ for $i\in\{2,3\}$.}
\label{fig:tripod}
\end{figure}
In Sec.~\ref{sec:group},
we present the procedure to reduce the group velocity in the second window. Finally, we summarize in Sec.~\ref{sec:summary}.
\section{Stationary-Atom Optical Susceptibility}
\label{sec:stationaryatom}
Consider the closed $\pitchfork$ atomic model scheme depicted in Fig.~\ref{fig:tripod}
with electronic level~$|4\rangle$ coupled to the lower levels~$\ket{1}$,~$\ket{2}$, and~$\ket{3}$ by three coherent fields,
namely the probe, coupling, and signal Rabi frequencies,
$\Omega_\text{p}$, $\Omega_\text{c}$,~$\Omega_\text{s}$, respectively. The $\ket{1}\leftrightarrow\ket{2}$, $\ket{1}\leftrightarrow\ket{3}$,  and $\ket{2}\leftrightarrow\ket{3}$ transitions are dipole forbidden.
The three fields are detuned from the electronic transition frequency~$\omega_{\imath\jmath}$ between states $|\imath\rangle$ and $|\jmath\rangle$ by
\begin{align}
\label{eq:deltap}
	\delta_\text{p}:=&\omega_{41}-\omega_\text{p},	\\
\label{eq:deltac}
	\delta_\text{c}:=&\omega_{42}-\omega_\text{c},\\
\label{eq:deltas}
	\delta_\text{s}:=&\omega_{43}-\omega_\text{p},
\end{align}
respectively.

The analytical steady-state density-matrix element ($\rho_{\imath\jmath}$) solution for a stationary atom,
to first order in the probe-field Rabi frequency,
can be approximated from the exact expression~\cite{Hessa2013} as
\begin{widetext}
\begin{equation}
\begin{split}
	\rho^{(1)}_{14}\approx\text{i}\Omega_\text{p}\frac{\left(\rho_{11}-\rho_{44}\right)(\Gamma_{43}+2\text{i}\delta_\text{s}+\frac{\vert\Omega_\text{c}\vert^2}{\Gamma_{32}+2\text{i}\delta_\text{sc}})+\left(\rho_{44}-\rho_{33}\right)\frac{\vert\Omega_\text{s}\vert^2}{\gamma_3-2\text{i}\delta_\text{ps}}}{(\Gamma_{43}+2\text{i}\delta_\text{s}+\frac{\vert\Omega_\text{c}\vert^2}{\Gamma_{32}+2\text{i}\delta_\text{sc}})(\gamma_4-2\text{i}\delta_\text{p}+\frac{\vert\Omega_\text{c}\vert^2}{\gamma_2-2\text{i}\delta_\text{pc}}+\frac{\vert\Omega_\text{s}\vert^2}{\gamma_{3}-2\text{i}\delta_\text{ps}})},
\end{split}
\label{eq:rho143}
\end{equation}
\end{widetext}
where
\begin{equation}
\label{eq:deltaxy}
	\delta_\text{xy}:=\delta_\text{x}-\delta_\text{y}
\end{equation}
is the two-photon detuning.
Details concerning the derivation of Eq.~(\ref{eq:rho143}) appear in Appendix~\ref{sec:stationary}.

We verified this expression numerically for weak signal and weaker probe Rabi frequencies, i.e.,
for the condition
\begin{equation}
\left|\Omega_\text{c}\right|^2\gg\left|\Omega_\text{s}\right|^2\gg|\Omega_\text{p}|^2.
\end{equation}
The decay rates in Eq.~(\ref{eq:rho143}) are
\begin{equation}
	\gamma_\jmath:=\sum_{\imath<\jmath}(\gamma_{\jmath\imath}+\gamma_{\phi\jmath}),
\end{equation}
and the coherence decay rate is
\begin{equation}
	\Gamma_\text{kl}=\gamma_\text{k}+\gamma_\text{l}.
\end{equation}
The dephasing rate between the forbidden transitions is not zero;
therefore, $\gamma_2=\gamma_{\phi 2}$ and $\gamma_3=\gamma_{\phi 3}$.

In our system, we impose the equal-population condition
\begin{equation}
\label{eq:equalpop}
	\rho_{11}\approx\rho_{33}\approx0.5.
\end{equation}
Condition~(\ref{eq:equalpop}) makes the equations approximately solvable analytically
as the equations of motion for population~(\ref{eq:Droh14}) are effectively decoupled from the equations of motion for coherence~(\ref{eq:Droh44}).
Condition~(\ref{eq:equalpop}) is achieved by incoherent excitation from
ground state~$\ket{1}$ to the excited state~$\ket{4}$ with constant pumping rate. Thus, the diagonal matrix elements are held constant by conditions~(\ref{eq:equalpop}). See Appendix~\ref{subsec:population} for details 
concerning the atomic population in our scheme.
The optical linear susceptibility for an atomic gas in three dimensions with~$\mathcal{N}$ the atomic density
and~$\bm{d}_{14}$ the dipole moment  is
\begin{equation}
\label{eq:chi1p}
	\chi^{(1)}_\text{p}
		=\eta\frac{\rho^{(1)}_{14}}{\Omega_\text{p}},
	\eta=\frac{\mathcal{N}\left|\bm{d}_{14}\right|^2}{\epsilon_0\hslash}.
\end{equation}
We can substitute Eq.~(\ref{eq:rho143})
into the numerator for~$\chi^{(1)}_\text{p}$ in Eq.~(\ref{eq:chi1p}),
which is complicated so we express~$\chi^{(1)}_\text{p}$ as
\begin{equation}
	\chi^{(1)}_\text{p}=\frac{\text{i}\eta}{2(B_1+2\text{i}B_2)}\left(1-\frac{C_1+2\text{i}C_2}{A_1-2\text{i}A_2}\right)
\label{eq:suscep}
\end{equation}
with the terms~$A_{1,2}$, $B_{1,2}$, and~$C_{1,2}$ explained in the following.

To simplify Eq.~(\ref{eq:rho143}),
we fix the value $\rho_{44}=0$. This is always true because the atoms are trapped to the dark state leaving level~$|4\rangle$ unpopulated. The population of the other three levels depends on the Rabi frequency of the driving fields.
See Appendix~\ref{subsec:population} for more details of the dark-state analysis and state populations.

The variables in Eq.~(\ref{eq:suscep}) are
\begin{align}
	A_1:=&\Gamma_{43}+\frac{\left|\Omega_\text{c}\right|^2\Gamma_{32}}{\Gamma_{32}^2+4\delta^2_\text{sc}},\;
	A_2:=\frac{\left|\Omega_\text{c}\right|^2\delta_\text{sc}}{\Gamma_{32}^2+4\delta^2_\text{sc}}-\delta_\text{s},
			\nonumber\\
	B_1:=&\gamma_4+\frac{\left|\Omega_\text{c}\right|^2\gamma_2}{\gamma_2^2
		+4\delta^2_\text{pc}}+\frac{\left|\Omega_\text{s}\right|^2\gamma_3}{\gamma^2_3+4\delta^2_\text{ps}},
			\nonumber\\
	B_2:=&\frac{\left|\Omega_\text{c}\right|^2\delta_\text{pc}}{\gamma_2^2
		+4\delta^2_\text{pc}}+\frac{\left|\Omega_\text{s}\right|^2\delta_\text{ps}}
		{\gamma^2_3+4\delta^2_\text{ps}}-\delta_\text{p},
			\\
	C_1:=&\frac{\left|\Omega_\text{s}\right|^2\gamma_3}{\gamma^2_3+4\delta^2_\text{ps}},
	C_2:=\frac{\left|\Omega_\text{s}\right|^2\delta_\text{ps}}{\gamma^2_3+4\delta^2_\text{ps}},\nonumber
\end{align}
We now have expressions for the steady-state solution~(\ref{eq:rho143})
and the corresponding susceptibilities for the probe field~(\ref{eq:suscep}).

\begin{figure}
\centering
 \subfloat[\label{ImChip}]{%
	 \includegraphics[width=0.4\textwidth]{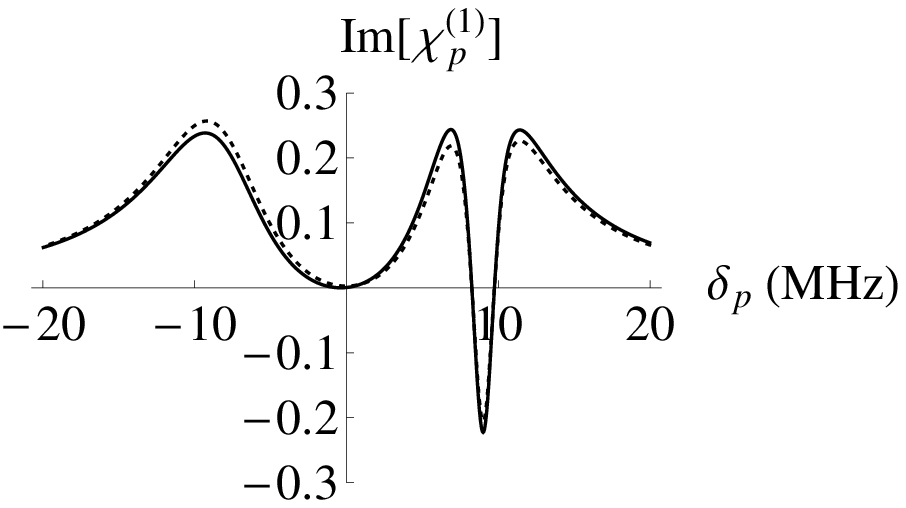}
 }
 \hfill
\subfloat[\label{ReChip}]{%
	 \includegraphics[width=0.4\textwidth]{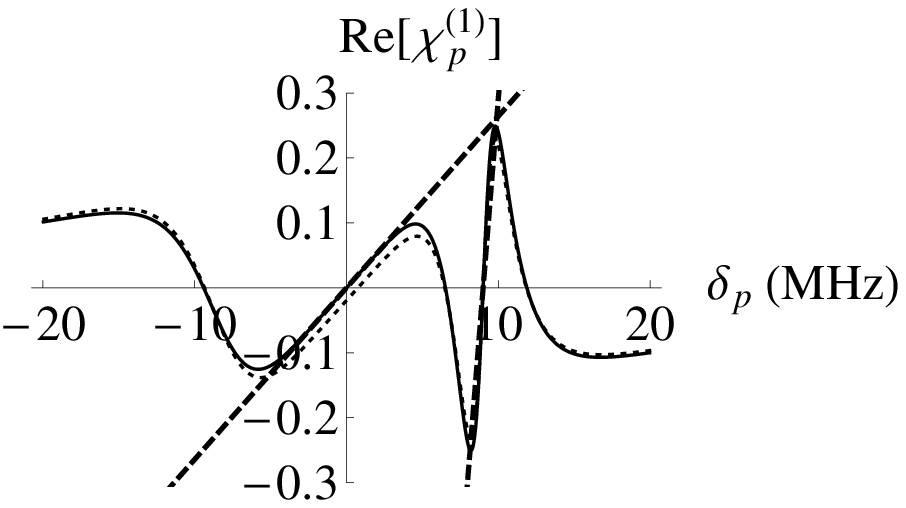}
}
	\caption{%
		(a)~Absorption  and~(b)~dispersion as a function of the probe detuning $\delta_\text{p}$, with  
		numerical (dotted line line), analytical (solid line),
		and approximate linear equation (dashed line line)
		for $\gamma_4=18$ MHz, $\gamma_3=10$ kHz, 
		$\gamma_2=40$ kHz, $\Omega_\text{c}=\gamma_4$, 
		$\Omega_\text{s}=0.3\gamma_4$, $\Omega_\text{p}=0.05\gamma_4$, $\delta_\text{s}=9$MHz, and $\delta_\text{c}=0$.%
		}
\label{fig:Re0k}
\end{figure}
Expression~(\ref{eq:suscep}) is used to calculate and plot the susceptibility,
whose imaginary part is shown in Fig.~\ref{fig:Re0k}(a), and whose real part is shown in Fig.~\ref{fig:Re0k}(b).
This absorption plot clearly displays the first probe window centered at~$\delta_\text{p}=\delta_\text{c}$
and the second EIT window centered at~$\delta_\text{p}=\delta_\text{s}\neq\delta_\text{c}$.
\subsection{Linewidth and Group Velocity}
\label{subsec:linewidthgroup}
The linewidth of each transparency window $\imath\in\{1,2\}$
is given by the half-width at half-maximum (HWHM)~$\daleth_\imath$.
The half-maximum values~$\varkappa_\imath$
are determined first by finding the maximum $h_{\text{max}_\imath}$
and the minimum $h_{\text{min}_\imath}$ values of  window $\imath$
as shown in Appendix~\ref{subsec:stationaryatom},
and then calculating
\begin{equation}
	\varkappa_\imath
		=\frac{h_{\text{max}_\imath}
				+h_{\text{min}_\imath}}{2}.	
\end{equation}
By solving
\begin{equation}
	\text{Im}[\chi^{(1)}_\text{p}] =\varkappa_\imath
\end{equation}
for $\delta_\text{pc}$ and  $\delta_\text{ps}$ separately,
$\daleth_1$ and $\daleth_2$ are determined respectively, 
with $\delta_\text{pc}=\daleth_1$ and  $\delta_\text{ps}=\daleth_2$.
In Appendix~\ref{subsec:stationaryatom} we also discuss the requirements
\begin{equation}
\label{eq:hom1}
	\left|\Omega_\text{c}\right|^2\gg\gamma_2\gamma_4
\end{equation}
and
\begin{equation}
\label{eq:hom2}
	\left|\Omega_\text{s}\right|^2\gg\gamma_3\gamma_4
\end{equation}
to overcome homogeneous broadening~\cite{Fleischhauer2005} at the first and second windows respectively, and to observe the presence of the EIT windows.

If conditions~(\ref{eq:hom1}) and~(\ref{eq:hom2})
are satisfied,
the HWHMs of the first and second window are 
\begin{equation}
	\daleth_1=\frac{\left|\Omega_\text{c}\right|^2}{\gamma_4+\sqrt{4\left|\Omega_\text{c}\right|^2+\gamma^2_4}},\;
	\daleth_2=\frac{\left|\Omega_\text{s}\right|^2}{2\sqrt{\gamma^2_4+\left|\Omega_\text{s}\right|^2}},
\label{eq:fw2}
\end{equation}
respectively.
Probe dispersion is shown in Fig.~\ref{fig:Re0k}(b). For detuning~$\delta_\text{p}$
chosen at the center of each window,
dispersion is zero or close to zero.

For each of windows~1 and~2,
group velocity is~\cite{Rebic2004}
\begin{equation}
\label{eq:vg}
	v_\text{g}
		\approx\frac{2c}{n_\text{g}},\;
	n_\text{g}=(\omega_0-\delta_\text{p})
		\left.\frac{\partial \text{Re}[\chi^{(1)}_\text{p}]}{\partial\delta_\text{p}}\right\vert_{\delta_\text{cen}}
\end{equation}
for~$n_\text{g}$ the group index,
~$\delta_\text{cen}$ the detuning at the center of each window (1 and~2),
and $\omega_0$ the transition frequency between levels~$\ket{1}$ and~$|4\rangle$.
Detuning~$\delta_\text{cen}$ equals~$\delta_\text{c}$ at the first window and equals~$\delta_\text{s}$ at the second window.
The partial derivative of the dispersion in the denominator is determined by
\begin{equation}
	\left.\frac{\partial \text{Re}[\chi^{(1)}_\text{p}]}{\partial\delta_\text{p}}\right\vert_{\delta_\text{cen}}
		=\lim_{\delta_\text{p}\to \delta_\text{cen}}
			\frac{\text{Re}[\chi^{(1)}_\text{p}(\delta_\text{p})]
				-\text{Re}[\chi^{(1)}_\text{p}(\delta_\text{p}=\delta_\text{cen})]}{\delta_\text{p}-\delta_\text{cen}}
\label{eq:slope}
\end{equation}
Therefore, the partial derivative of the dispersion 
\begin{equation}
	\left.\frac{\partial \text{Re}[\chi^{(1)}_\text{p}]}{\partial\delta_\text{p}}\right\vert_{\delta_\text{c}}=\frac{\eta\left|\Omega_\text{c}\right|^2}{\left(\gamma_2\gamma_4+\left|\Omega_\text{c}\right|^2\right)^2}
\label{l1}
\end{equation}
at the center of the first window and 
\begin{equation}
	\left.\frac{\partial \text{Re}[\chi^{(1)}_\text{p}]}{\partial\delta_\text{p}}\right\vert_{\delta_\text{s}}=\frac{\eta\left|\Omega_\text{s}\right|^2}{(\gamma_3\gamma_4+\left|\Omega_\text{s}\right|^2)^2}
\label{l2}
\end{equation}
at the center of the second window.
Equations.~(\ref{l1}) and~(\ref{l2}) yield the slope of the tangent line to points
$\delta_\text{p}=\delta_\text{cen}$ as shown in Fig.~\ref{fig:Re0k}(b).

In Fig.~\ref{fig:Re0k}(b),
the group velocity is shown to be approximately constant in each of the two EIT windows,
which can be seen by the straight-line tangents.
The group velocity scales inversely with slope
so the ratio of group velocities for each EIT window is the inverse of the 
ratio of the slopes for each window.
From Eqs.~(\ref{l1}) and~(\ref{l2}) and from Fig.~\ref{fig:Re0k}(b),
the group velocity at the first window evidently exceeds the group velocity at the second window for the given parameters.\

Under conditions~(\ref{eq:hom1}) and~(\ref{eq:hom2}),
the group velocity reduces to
\begin{equation}
\label{eq:groupvelocity1}
	v_{\text{g}_1}=\frac{2c}{\eta}\frac{\left|\Omega_\text{c}\right|^2}{\omega_{14}}
\end{equation}
at the first window
and to
\begin{equation}
\label{eq:groupvelocity2}
	v_{\text{g}_2}=\frac{2c}{\eta}\frac{\left|\Omega_\text{s}\right|^2}{\omega_{34}}
\end{equation}
at the second window.
Hence, for stationary atoms, the group velocities in both windows are linearly proportional to the intensities of the respective driving fields.

\section{Doppler-broadened optical susceptibility}
\label{sec:Doppler}

At  non-zero temperature atoms move randomly due to thermal energy.
Thermal atomic motion leads to a spreading of the absorbed frequency due to the Doppler effect, which  broadens the optical line profile and is known as Doppler broadening.

In this section we solve susceptibility numerically and also derive approximate analytical expressions
as a function of temperature.
These results are used to find the widths of transparency windows and also group velocities of the
probe field in each of the two DDEIT windows.
Our approximate analytical technique is based on approximating the Maxwell-Boltzmann velocity
distribution for atoms by Lorentzian distributions over the narrow but relevant domain of small atomic velocities~\cite{Javan2002}.
This approximation is valid as large velocities are sufficiently detuned so as not to affect the optics.

In our scheme, the electromagnetic field passes through a gas of atoms at temperature~$T$.
Each atom of mass~$m$ has a velocity~$\bm v$ obeying the Gaussian Maxwell-Boltzmann distribution
\begin{equation}
\label{eq:fv}
	f(v)=\frac{1}{u\sqrt{\pi}}\exp\left(-\frac{v^2}{u^2}\right),\;
	u=\sqrt{\frac{2k T}{m}}
\end{equation}
with~$v$ the component of velocity~$\bm v$ in the direction of the three co-propagating
signal, probe, and coupling fields.

One effect of moving atoms is detuning of resonant frequencies due to the Doppler shift,
which results in a velocity-dependent probe-field susceptibility~$\chi_\text{p}(v)$.
For our Doppler-broadened system,
the susceptibility is thus averaged over the entire velocity distribution
according to~\cite{Julio1995}
\begin{equation}
	\bar{\chi}_\text{p}:=\int^{\infty}_{-\infty}{\chi_\text{p}(v) f(v) \text{d}v}.
\label{eq:Dopp}
\end{equation}
The velocity-dependent expression for susceptibility is obtained from Eq.~(\ref{eq:suscep})
by the replacement
\begin{equation}
	\delta_\text{x}\mapsto\delta_\text{x}+\frac{v\omega_\text{x}}{c},\;
	\text{x}\in\{\text{p,c,s}\},
\end{equation}
for
\begin{equation}
\label{eq:omegax}
	\omega_\text{x}
		=\left\{
			\begin{array}{lr}
			\omega_{14}\equiv\omega_0,&\text{x}=\text{p},\\
			\omega_{24},&\text{x}=\text{c},\\
			\omega_{34},&\text{x}=\text{s},
			\end{array}
		\right.
\end{equation}
the atomic frequencies and~$c$ the speed of light $in \text{ }vacuo$.

Our scheme relies on neglecting Doppler effect on two-photon detuning $\delta_\text{xy}$~(\ref{eq:deltaxy}),
which is achieved for the co-propagating fields driving 
approximately equal transition frequencies:
\begin{equation}
	\omega_0\equiv\omega_{14}\approx\omega_{24}\approx\omega_{34}.
\end{equation}
This choice is commensurate with our case of a~$^{87}$Rb gas.
For this atom,
we assign~$\ket{1}$,~$\ket{2}$, and~$\ket{3}$ to the~$5S_{1/2}$ level
with~$F=1$,~$m_F=0$ and~$F=2$,~$m_F=\{-2,0\}$ respectively.
Level~$|4\rangle$ corresponds to the~$5P_{1/2}$ level with~$F=2$ and~$m_F=-1$.
Therefore,  the quantities~$\{\delta_\text{xy}\}$ in Eqs.~(\ref{eq:suscep})
do not change under Doppler broadening.

Integration of Eq.~(\ref{eq:Dopp}) corresponds to a convolution of Lorentzian~$\chi_\text{p}$
with the Gaussian profile,
which is known as the Voigt profile~\cite{Pag08}. The Voigt profile can be solved numerically
but is hard to solve analytically~\cite{Julio1995,VA96,Ye2002,Li2007}.

The lack of an exact analytical solution inhibits finding a simple expression relating
the group velocity or width of each EIT window to Doppler broadening.
Instead, we approximate the Maxwell-Boltzmann distribution by a Lorentzian function over a narrow velocity domain~\cite{Javan2002} to obtain an approximate analytical expression for the optical susceptibility.
This approximation is valid insofar as we are interested in the optical response near the spectral center.

\subsection{Lorentzian line-shape approximation}
\label{sec:Lor}
In this subsection, we determine an analytical approximation to the optical susceptibility
for a Doppler-broadened system.
Our approximation uses a Lorentzian fit to the Maxwell-Boltzmann velocity distribution over a narrow range of velocity.
We use this approximation to show that the first probe transparency window
is independent of the signal-field Rabi frequency and the second transparency window is nonlinear in the signal-field Rabi frequency. Furthermore, we derive the connection between the transparency window and the Doppler broadening width, which is directly dependent on the temperature.

The Lorentzian line-shape function~\cite{Yuri2002}
\begin{equation}
	L\left(\frac{v\omega_0}{c}\right)=\frac{1}{\sqrt{\pi}}\frac{W_\text{L}}{W^2_\text{L}+\left(\frac{v\omega_0}{c}\right)^2}
\label{Lore}
\end{equation}
is a function of the atomic velocity
with~$W_\text{L}$ is the HWHM of the Lorentzian profile.
To see that the Lorentzian~(\ref{Lore}) approximates the Gaussian~(\ref{eq:fv}) well over
a narrow domain, we first write both functions as Maclaurin series.
The Gaussian~(\ref{eq:fv}) is approximated by
\begin{align}
\label{eq:fvapprox}
	f\left(\frac{v\omega_0}{c}\right)=&\frac{\sqrt{\ln{2}}}{\sqrt{\pi}W_\text{G}}-\frac{\omega^2_0(\sqrt{\ln{2}})^3}{c^2\sqrt{\pi}W^3_\text{G}}v^2\nonumber\\&
+\frac{\omega^4_0(\sqrt{\ln{2}})^5}{c^4\sqrt{\pi}W^4_\text{G}}v^5-\cdots
\end{align}
 with 
\begin{equation}
\label{eq:WG}
	W_\text{G}:=\frac{\omega_0}{c}\sqrt{\frac{2k T \ln2}{m}}
\end{equation}
the HWHM of the Gaussian profile
and 
\begin{align}
	L\left(\frac{v\omega_0}{c}\right)=&\frac{1}{\sqrt{\pi}W_\text{L}}-\frac{\omega^2_0}{c^2\sqrt{\pi}W^3_\text{L}}v^2\nonumber\\&
+\frac{\omega^4_0}{c^4\sqrt{\pi}W^5_\text{L}}v^5-\cdots,
\label{expaLor}
\end{align}
for
\begin{equation}
	-1<\frac{\omega_0v}{c W_\text{L}}<1.
\end{equation}

\begin{figure}
\includegraphics[width=0.8\columnwidth]{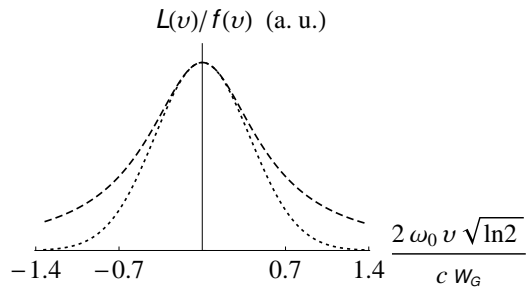}
	\caption{
		Plot of Lorentzian function (dashed line) and Gaussian function (dotted line) vs. normalized atomic velocity.}
\label{LoGa}
\end{figure}

The two expansions~(\ref{eq:fvapprox}) and~(\ref{expaLor})
are approximately equal under the conditions that 
\begin{equation}
\label{eq:WLWG}
	W_\text{L}=\frac{1}{\sqrt{\ln2}}W_\text{G},
\end{equation}
for
\begin{equation}
	-1\ll\frac{\omega_0v}{c W_\text{G}}\sqrt{\ln{2}}\ll1.
\end{equation}
Combining Eqs.~(\ref{eq:WG}) and~(\ref{eq:WLWG}) yields the connection between the Lorentzian linewidth and the temperature.
These conditions are satisfied near the center of both function profiles as shown in Fig.~\ref{LoGa}, where the higher-order terms of Eqs.~(\ref{eq:fvapprox}) and~(\ref{expaLor}) have insignificant influence.

Integration of Eq.~(\ref{eq:Dopp}) using $L(v)$ instead of~$f(v)$  has two terms evaluated with the contour integral using the residue theorem.
The final optical susceptibility,
including the Doppler broadening effect, is
\begin{equation}
	\bar{\chi}_\text{p}(\delta_\text{p})
%		=\chi^{(1)}+\chi^{(3)}\left|\Omega_\text{s}\right|^2
		=I_1(\delta_\text{p})+I_2(\delta_\text{p})
\label{eq:susplorent}
\end{equation}
with~$\delta_\text{p}$ the detuning~(\ref{eq:deltap}).
The terms on the right-hand side of Eq.~(\ref{eq:susplorent}) are
\begin{equation}
\label{eq:I1}
	I_1=\frac{\text{i}\eta}{2}\frac{\sqrt{\pi }}{B_1+2\text{i}B_2+W_\text{L}}
\end{equation}
and
\begin{align}
	I_2
		=&-\frac{\text{i}\eta}{2}\frac{\sqrt{\pi }}{B_1
			+W_L+2\text{i}B_2}\frac{C_1+\text{i}C_2}{A_1-W_L-2\text{i}A_2}
					\nonumber\\&
		-\frac{C_1+\text{i}C_2}{A_1+B_1+2\text{i}(B_2-A_2)}
				\nonumber\\&\times
		\frac{\text{i}\eta W_\text{L}\sqrt{\pi }}{W^2_L+4A^2_2-A^2_1+4\text{i}A_1A_2}.
\label{eq:I2}
\end{align}

The HWHM~$\bar{\daleth}_1$ of the first transparency window,
and the group velocity for this window,
depend on~$I_1(\delta_\text{p})$
but not on~$I_2(\delta_\text{p})$ over the domain of~$\delta_\text{p}$ pertaining to the first window.
In the case of the second transparency window for the probe field,
both $I_1(\delta_\text{p})$ and $I_2(\delta_\text{p})$ are non negligible 
for calculating the HWHM~$\bar{\daleth}_2$ and group velocity.

\begin{figure}
\centering
    \subfloat[\label{subfig-absorption1}]{%
      \includegraphics[width=0.5\columnwidth]{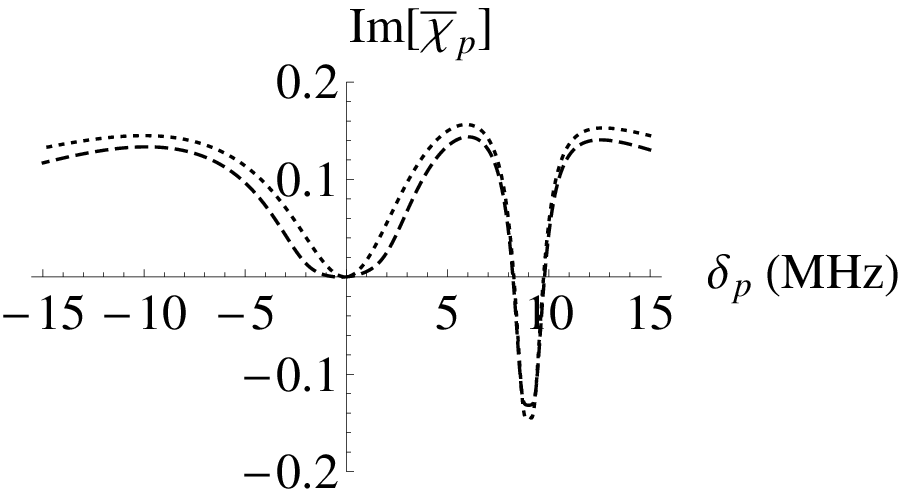}
    }
    \subfloat[\label{subfig-dispersion1}]{%
       \includegraphics[width=0.5\columnwidth]{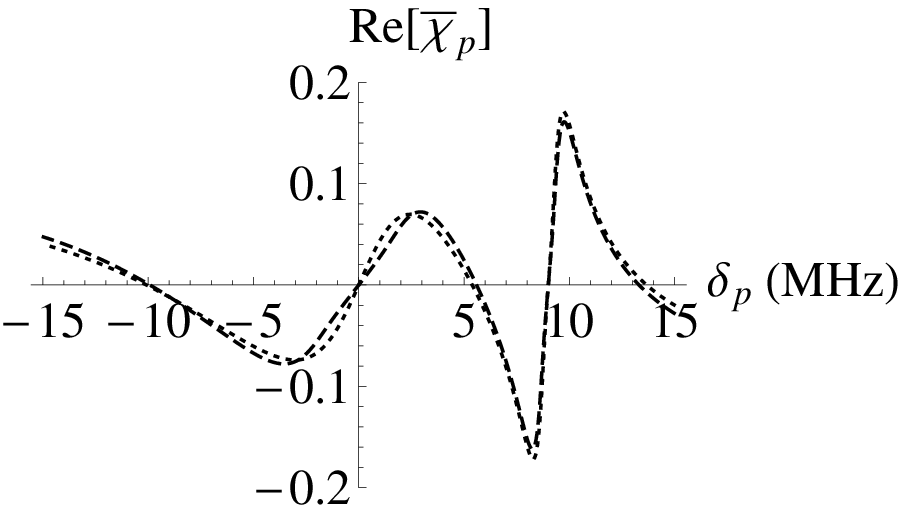}
    }
 \hfill
\subfloat[\label{subfig-absorption100}]{%
      \includegraphics[width=0.5\columnwidth]{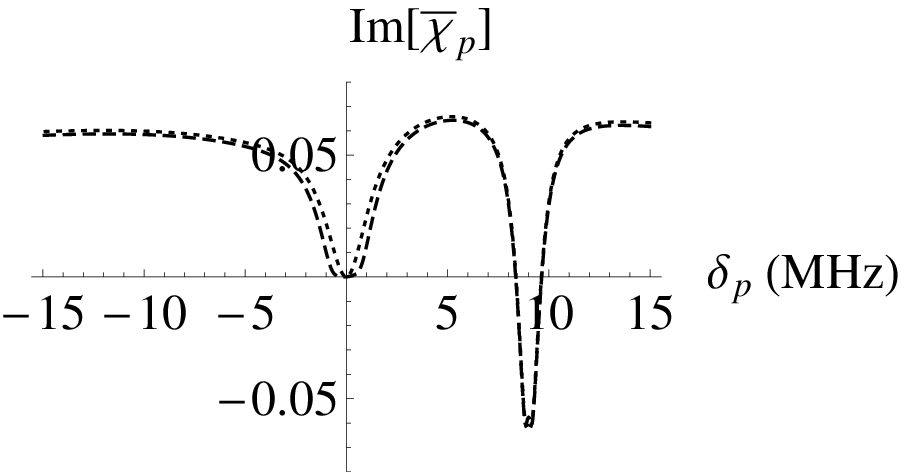}
    }
    \subfloat[\label{subfig-dispersion100}]{%
       \includegraphics[width=0.5\columnwidth]{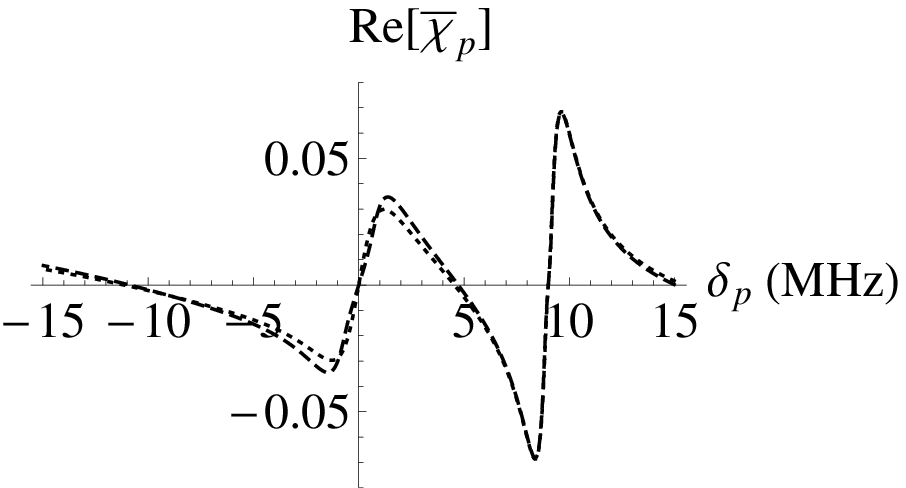}
    }
 \hfill
\subfloat[\label{subfig-absorption1000}]{%
      \includegraphics[width=0.48\columnwidth]{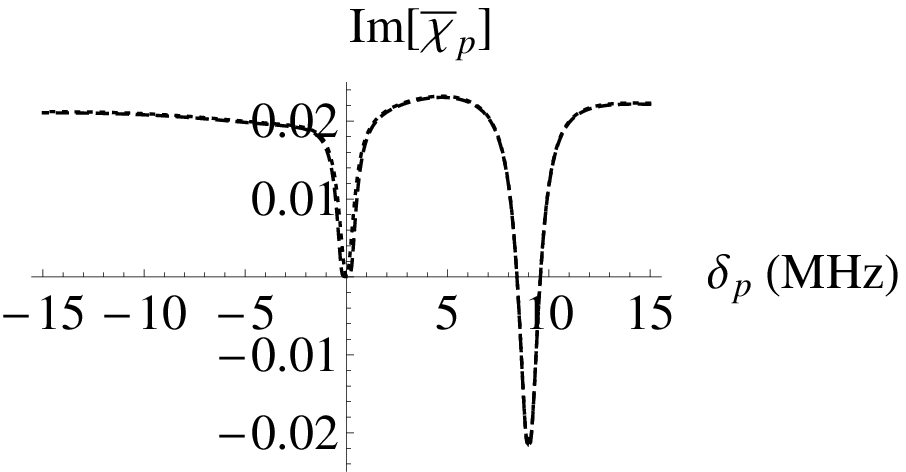}
    }
   \hfill
    \subfloat[\label{subfig-dispersion1000}]{%
       \includegraphics[width=0.48\columnwidth]{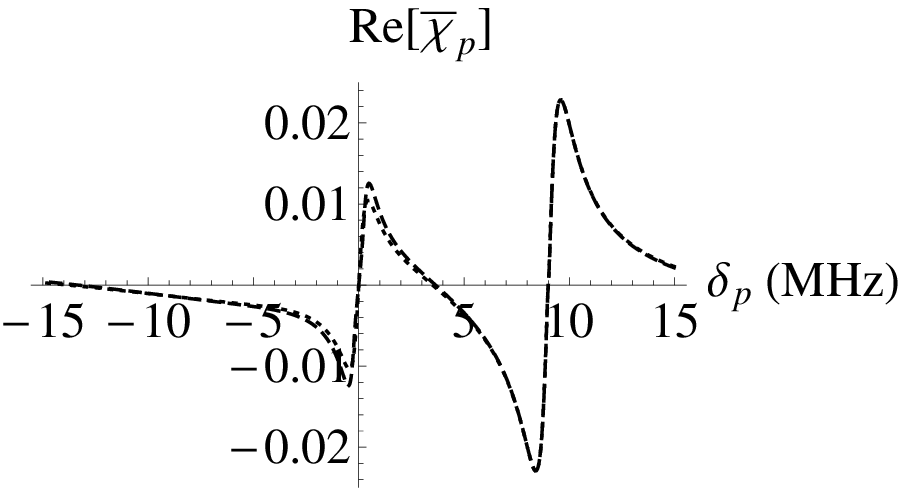}
    }
    \caption{%
    	Plots of Im$[\chi^{(1)}_\text{p}]$ and Re$[\chi^{(1)}_\text{p}]$ 
		vs probe detuning  $\delta_\text{p}$ at different temperatures
		for $\gamma_4=18 $MHz, $\gamma_3=10$ kHz, 
		$\gamma_2=40 $kHz, $\Omega_\text{c}=\gamma_4$, 
		$\Omega_\text{s}=0.35\gamma_4$, $\delta_s=9$MHz, and $\delta_c=0$. (a), (c) and (e) are  Im$[\bar{\chi}_\text{p}]$ and~(b), (d), and (f) are  Re$[\bar{\chi}_\text{p}]$. We set $T$= (1, 10, 100) K  for (a),(b), (c),(d) and (e),(f) respectively, which is equivalent to $W_\text{L}=(34.8, 110, 348) $MHz, respectively. The dotted line-line corresponds to the analytical solution using the Lorentzian line-shape function, whereas the dashed line line is the numerical solution using the Maxwell-Boltzmann distribution function.}
 \label{fig:LGd}
 \end{figure}

In Fig.~\ref{fig:LGd}, we plot the imaginary and real terms of the susceptibility~$\chi_\text{p}^{(1)}$
as a function of the probe-field detuning~$\delta_\text{p}$
at various temperature values based on the average susceptibility~(\ref{eq:Dopp})
for the Maxwell-Boltzmann distribution function~$f(v)$ and for the approximation
using the Lorentzian function~$L(v)$.
At Low temperatures,
for which the broadening is low,
there is a discrepancy between the two functions. 

At higher temperatures,
for which
\begin{equation}
\label{eq:ValAppro}
	W^2_\text{L}\gg\gamma^2_4,
\end{equation}
the numerical data agree with the analytical data near the center as seen by comparing the two plots.
The plots differ at the tail,
which describes far-off-resonant atoms whose contribution is negligible.
This numerical result validates the Lorentzian approximation for condition~(\ref{eq:ValAppro}) near the center, which leads to a rather simple form of the inhomogeneously broadened susceptibility.

Analyzing the numerical result reveals that condition~(\ref{eq:hom1})
and condition
\begin{equation}
\label{eq:inhom1}
	\left|\Omega_\text{c}\right|^2>\gamma_2W_\text{L}
\end{equation}
are required to observe the first transparency window.
These conditions~(\ref{eq:hom1}) and~(\ref{eq:inhom1}) eliminate the homogeneous broadening and reduce the effect of inhomogeneous broadening, respectively.
At a temperature for which the Doppler broadening satisfies condition~(\ref{eq:ValAppro}), and Eqs.~(\ref{eq:susplorent})-(\ref{eq:I2}) are a valid approximation,
satisfying condition~(\ref{eq:inhom1}) certainly implies satisfying  condition~(\ref{eq:hom1}).

As shown in Fig.~\ref{fig:LGd},
the width~$\daleth_2$ of the second transparency window is not noticeably affected
by varying the Doppler width~$W_\text{L}$.
The reason for the robustness of~$\daleth_2$ is that the nonlinear interaction in~$I_2$,
but not in~$I_1$,
protects the second window from deleterious temperature effects.
Therefore, the strong-signal-field condition is not required to overcome Doppler broadening damaging the second transparency window.
In other words, condition
\begin{equation}
\label{eq:inhom2}
	\left|\Omega_\text{s}\right|^2>\gamma_3 W_\text{L}
\end{equation}
is no longer mandatory to observe the second window. 

Condition~(\ref{eq:hom2})
is still required to eliminate the homogeneous broadening for significant transparency at the second window; see Appendix~\ref{subsec:stationaryatom} for detailed mathematical proofs of the conditions
required to observe the transparency windows.
Furthermore, the relaxation of condition~(\ref{eq:inhom2}) leads to further reduction
of group velocity in Doppler-broadened media, which was limited by the Doppler width
appearing in the right-hand side of condition~(\ref{eq:inhom2}).

The two terms $\gamma_2W_\text{L}$ in Eq.~(\ref{eq:inhom1})
and $\gamma_3W_\text{L}$ in Eq.~(\ref{eq:inhom2})
quantify the inhomogeneous broadening of the two EIT windows.
In other words, the Doppler broadening alone is not the whole story;
rather, the products~$\gamma_{2,3}W_\text{L}$
incorporating the rates~$\gamma_2$ and~$\gamma_3$ are the key quantities.
In Sec.~\ref{subsec:width}
we derive the linewidth and the group velocity
for which the requisite conditions~(\ref{eq:hom1}) and~(\ref{eq:hom2})
for eliminating homogeneous broadening, are always satisfied for both windows.

\subsection{Width of the transparency windows}
\label{subsec:width}

The EIT width in a three-level Doppler-broadened~$\Lambda$ system can be maintained by keeping the temperature of the system constant while changing the driving field~\cite{Taichenachev2000, Yuri2002, Javan2002, Ye2002, Figueroa2006}.
Here, we follow a different approach by studying the dependence of the linewidth on temperature while fixing the intensity of the driving fields. The intensities of the driving fields are chosen such to eliminate the homogeneous broadening.\

\begin{figure} 
\setbox1=\hbox{3cm}{\includegraphics[width=0.95\columnwidth]{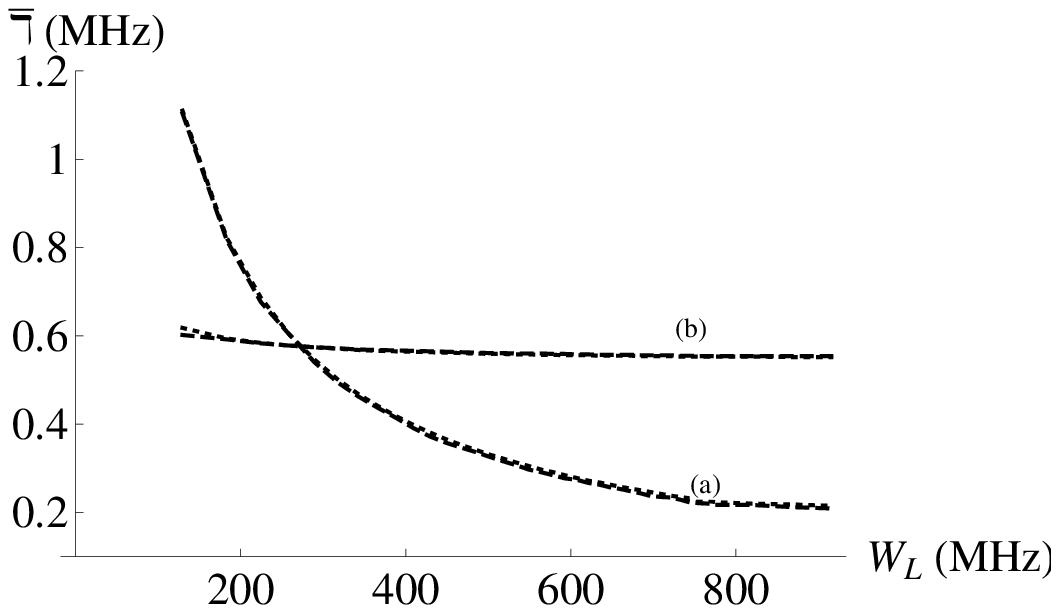}}\llap{\raisebox{2.6cm}{\includegraphics[height=2.5cm]{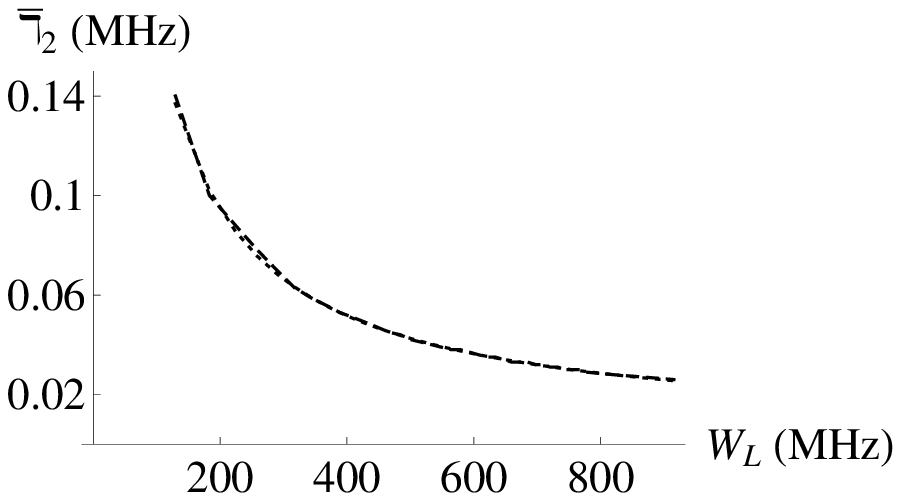}}}
\caption{%
	Numerical (dashed line) and analytical (dotted line) solutions of the HWHM ($\bar\daleth$)
	for the (a)~first and~(b)~second EIT transparency windows vs Doppler width~$W_\text{L}$
	for $\gamma_4=18$ MHz, $\gamma_3=10$ kHz, 
	$\gamma_2=40$ kHz, $\Omega_\text{c}=\gamma_4$, 
	$\Omega_\text{s}=0.35\gamma_4$, $\delta_s=9 $MHz, and $\delta_c=0$.
	Inset:
	Numerical (dashed line) and analytical (dotted line) 
	HWHM of the second EIT window evaluated for the gain term eliminated.%
	}
\label{fig:Wtemp}
\end{figure}

The HWHM of the first window for the Doppler-broadened system is equal to
\begin{widetext}
\begin{equation}
	\bar{\daleth}_1
		=\frac{\left|\Omega_\text{c}\right|^2}{2}
			\Bigg[\frac{(2\gamma_2W_\text{L}+\left|\Omega_\text{c}\right|^2)}{2(\gamma_4+W_\text{L})^2(\gamma_2W_\text{L}+\left|\Omega_\text{c}\right|^2)-W_\text{L}(W_\text{L}+2\gamma_4)(2\gamma_2W_\text{L}+\left|\Omega_\text{c}\right|^2)}\Bigg]^{1/2}.
\label{Gamma1}
\end{equation}
\end{widetext}
The width decreases nonlinearly as the Doppler width~$W_\text{L}$ increases 
as shown in Fig.~\ref{fig:Wtemp}.
The condition $\left|\Omega_\text{c}\right|^2\gg\gamma_2W_\text{L}$ is valid for all~$W_\text{L}$ values in the figure.
For a high-intensity coupling field~(\ref{eq:inhom1}), the width of the first window reduces to
\begin{equation}
\bar{\daleth}_1=\frac{\left|\Omega_\text{c}\right|^2}{2\sqrt{W_\text{L}(2\gamma_4+W_\text{L})}}.
\end{equation}
The formula for HWHM can be further simplified if $W_\text{L}\gg\gamma_4$,
thereby yielding
\begin{equation}
\label{eq:width1}
	\bar{\daleth}_1=\frac{\left|\Omega_\text{c}\right|^2}{2W_\text{L}}. 
\end{equation}
This result is consistent with the previous result for a three-level~$\Lambda$ atom, 
subject to a high-intensity driving field,
for which the linewidth is proportional to the intensity of the driving field and inversely proportional to the Doppler width~\cite{Javan2002}.

The HWHM of the second window of the Doppler-broadened system
has a more complicated form than for the first window:
\begin{equation}
	\begin{split}
	\bar{\daleth}_2=\frac{\left|\Omega_\text{s}\right|^2}{2}\sqrt{\frac{(\gamma_4+W_\text{L})+W_\text{L}\left[\bar{\varkappa}_2(\gamma_4+W_\text{L})-\frac{1}{2}\right]}{4W_\text{L}\gamma^2_4\left[\frac{1}{2}-\bar{\varkappa}_2(\gamma_4+W_\text{L})\right]+\left|\Omega_\text{s}\right|^2(\gamma_4+W_\text{L})}}\\
\label{eq:Gamma2}
\end{split}
\end{equation}
where
\begin{align}
	\bar{\varkappa}_2
		=&\frac{\eta\sqrt{\pi}}{4}
			\Bigg\{\frac{2\gamma_3W_\text{L}+\Omega^2_\text{s}}{(\gamma_4+W_L)(\gamma_3W_\text{L}+\Omega^2_\text{s})}
					\nonumber\\&
			+\frac{2\gamma_3 W_\text{L}(\gamma_4-W_\text{L})+\Omega^2_\text{s}(2\gamma_4-W_\text{L})}
			{W_\text{L}\left[\gamma_3 W^2_\text{L}+(W_\text{L}-\gamma_4)\Omega^2_\text{s}\right]}
			\Bigg\}
\label{vaarkappa2}
\end{align}
is the half-maximum value of~Im$\bar{\chi}_{\text p}$ of the second window.
The dependence of the HWHM of the second window on Doppler width is shown in Fig.~\ref{fig:Wtemp}. The width of the second window slightly decreases as the Doppler width increases.

For large Doppler broadening,
$W_L\gg\gamma_4$,
$\bar{\varkappa}_2$ depends on the population difference $\rho_{11}-\rho_{33}$.
As we set $\rho_{11}\approx\rho_{33}\approx0.5$, $\bar{\varkappa}_2$
is always located at Im$\bar{\chi}_\text{p}\approx0$,
i.\ e., where absorption vanishes.
Consequently, the width of the second window remains approximately constant
with respect to Doppler width
\begin{equation}
	\bar{\daleth}_2=\frac{\left|\Omega_\text{s}\right|^2}{2\sqrt{2}\sqrt{\gamma^2_4+2\left|\Omega_\text{s}\right|^2}}.
\label{eq:width2}
\end{equation}
This independence Doppler broadening width response of the second window is
due to the gain described by Im$I_2$ of Eq.~(\ref{eq:susplorent}).
Expression~(\ref{eq:width2}) reveals that further reduction of the group velocity can be achieved by reducing the intensity of the signal field without losing the transparency window due to Doppler broadening.

The two EIT windows have the same width at the intercept point between the two curves 
as shown in Fig.~\ref{fig:Wtemp}.
For all values of Doppler width the signal field has lower intensity than the coupling field. The inset to Fig.~\ref{fig:Wtemp} shows how the second window would behave as a function of
$W_\text{L}$ if the nonlinear contribution~$I_2$ were suppressed.
This inset makes clear how important the optical nonlinearity is for achieving quite different 
temperature sensitivities of the two transparency windows for the probe field.
Mathematically, an effect of forcing $I_2\equiv 0$ is that the HWHM of the second transparency 
window is given by a modification of the HWHM of the first window~(\ref{Gamma1})
with the proviso that~$\Omega_\text{c}$ is replaced by~$\Omega_\text{s}$
and~$\gamma_2$ is replaced by~$\gamma_3$.

For atoms copropagating with the probe field, the gain term suppresses the narrowing  of the width results from Doppler broadening.
Generalizing the choice of atomic propagation direction relative to the direction of
the three driving fields
would of course lead to different results~\cite{VA96}.

In summary, Eq.~(\ref{eq:Gamma2}) is the full expression of the HWHM
of the second transparency window and accounts for the nonlinear interaction between the probe and the signal field.
Its behavior is depicted in Fig.~\ref{fig:Wtemp} and 
shows the insensitivity of the second transparency window on 
temperature,
which is represented by width~$W_\text{L}$.
Contrariwise, the first window is sensitive to~$W_\text{L}$.

\subsection{Group velocities at the transparency windows}
\label{sec:Gvel}
\begin{figure}
\centering
\setbox1=\hbox{5cm}{\includegraphics[width=1\columnwidth]{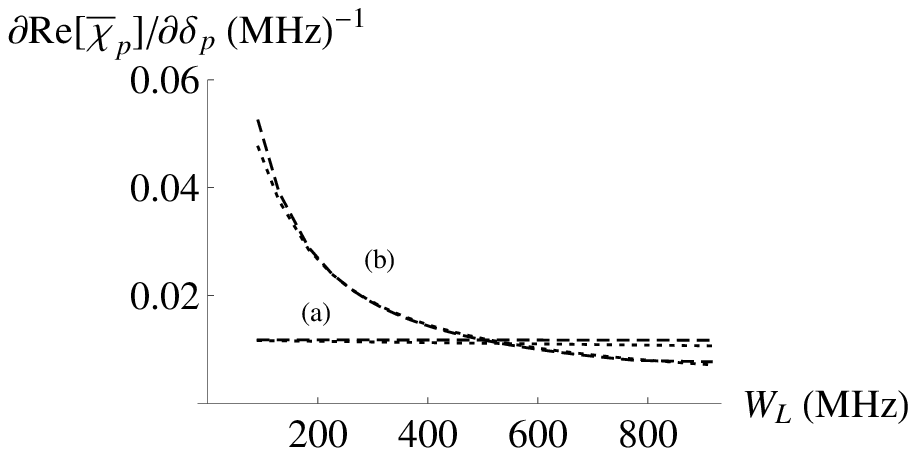}}\llap{\raisebox{2.5cm}{\includegraphics[height=2.3cm]{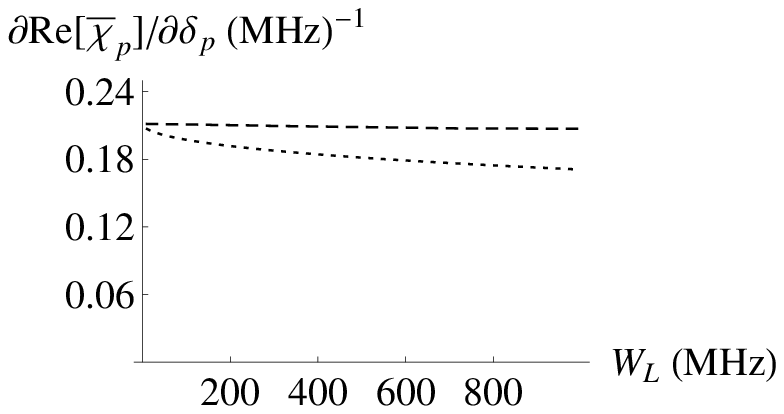}}}
\caption{%
	Plots of the numerical (dashed line) and analytical (dotted) results for the
	partial derivative of dispersion with respect to Doppler width for
	(a)~the first window and~(b)~the second window
	for $\gamma_4=18$ MHz, $\gamma_3=10$ kHz, 
	$\gamma_2=40$ kHz, $\Omega_\text{c}=1.5\gamma_4$, 
	$\Omega_\text{s}=0.5\gamma_4$, $\delta_s=13.5$ MHz, and $\delta_c=0$.\
	Inset: numerical (dashed line) and analytical (dotted line) results
	for the partial derivative of dispersion vs Doppler width at the second window for $I_2\equiv 0$.%
 	}
\label{fig:Dtemp}
\end{figure}

From Sec.~\ref{sec:Lor}, we have approximate analytical expressions for susceptibilities at the two transparency windows.
In this subsection we determine the derivative of the susceptibility with respect to the detuning~$\delta_\text{p}$ and use these partial derivatives of dispersion~(\ref{eq:slope})
to calculate the group velocities for the probe field in each of the two transparency windows.
The response of the partial derivative of dispersion with respect to Doppler broadening system
is shown in Fig.~\ref{fig:Dtemp}.
In this figure, numerical calculations show constant group velocity at the first window and a sharply increased group velocity at the second window.

The analytical expression for the group velocity of the Doppler-broadened system is evaluated using Eq.~(\ref{eq:vg}) but with the Doppler-broadened susceptibility~(\ref{eq:susplorent}) replaced the free Doppler-broadened susceptibility $\chi^{(1)}_\text{p}$~\cite{Kash1999}.
The partial derivative of Re$[\bar{\chi}_\text{p}]$ at the center of the first window is
\begin{equation}
\label{eq:DLF}
	\left.\frac{\partial \text{Re}[\bar{\chi}_\text{p}]}{\partial\delta_\text{p}}\right|_{\delta_\text{c}}
		=\frac{\eta\sqrt{\pi}|\Omega_\text{c}|^2}
			{\left(\gamma_2W_\text{L}+{\left|\Omega_\text{c}\right|^2}\right)^2}
\end{equation}
and at the center of the second window is
\begin{align}
	\left.\frac{\partial \text{Re}[\bar{\chi}_\text{p}]}{\partial\delta_\text{p}}\right|_{\delta_\text{s}}
		=&\frac{2\eta\sqrt{\pi}\left|\Omega_\text{s}\right|^2\gamma_4}{\left(\gamma_4-W_\text{L}\right)
			\left(\gamma_3W_\text{L}+\left|\Omega_\text{s}\right|^2\right)^2}
							\nonumber\\&
+\frac{4\eta\sqrt{\pi}\left|\Omega_\text{s}\right|^2\gamma_4}{W_\text{L}\left|\Omega_\text{s}\right|^4}.
\label{eq:DLS2}
\end{align}
For the first transparency window and for a strong coupling field~(\ref{eq:inhom1}),
the group velocity of the probe field at the center of the first window
has the same group velocity as for the Doppler-free case~(\ref{eq:groupvelocity1}).
The negligibility of the Doppler broadening effect is due to the intensity of the coupling field being large,  as can be explained from the analytical expression~(\ref{eq:DLF}).

\begin{figure} 
\includegraphics[width=1\columnwidth]{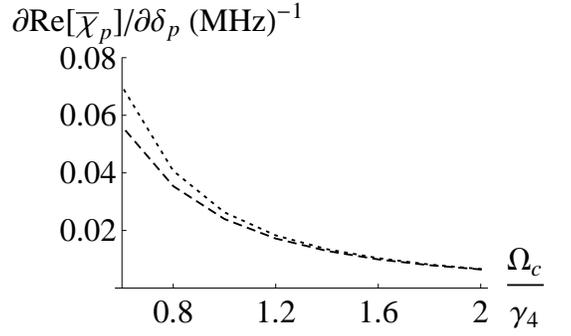}
\caption{%
	Plot of the numerical (dotted) and analytical (dashed line) results for the partial derivative of dispersion
	vs coupling field at the first window for Doppler width $W_\text{L}=409$ MHz, 
	$\gamma_4=18$ MHz, $\gamma_3=10$ kHz, $\gamma_2=40$ kHz.
	}
\label{DerCoup}
\end{figure}
Figure~\ref{fig:Dtemp} shows 
agreement between the analytical expression~(\ref{eq:DLF}) and the full numerical result 
applicable for small~$W_\text{L}$.
This agreement diminishes slightly as~$W_\text{L}$ increases.
Therefore, the Lorentzian function can be used to study
the Doppler-broadened dispersion response of the~$\Lambda$ configuration comprising the three states~$\ket{1}$,~$\ket{2}$, and~$|4\rangle$
provided that condition~(\ref{eq:inhom1}) is satisfied.

Our analytical expression is reliable in practical parameter regimes.
This agreement between the analytical Lorentzian approximation 
and the full numerical result under condition~(\ref{eq:inhom1})
is presented in Fig.~\ref{DerCoup} for varying coupling-field Rabi frequency.

We establish reliability of our approximation by comparing to an approximate
Lorentzian expression derived for a~$\Lambda$ EIT system~\cite{Kash1999}.
In our notation,
their result for group index is
\begin{equation}
\label{eq:kash}
	n_\text{g}\propto\frac{\gamma_4\left|\Omega_\text{c}\right|^2}
		{\left[\gamma_2\left(\gamma_4+W_\text{L}\right)
			+\left|\Omega_\text{c}\right|^2\right]^2}
\end{equation}
with the relation between group index and derivative of dispersion (\ref{eq:DLF})
given by~Eq.~(\ref{eq:vg}).
We can neglect~$\gamma_4$ from~(\ref{eq:kash})
according to the approximation~(\ref{eq:hom1}).
Although result~(\ref{eq:kash})
is derived for a~$\Lambda$ system and our result~(\ref{eq:DLF}) for a~$\pitchfork$ system,
both results pertain to an EIT window in a strong-coupling regime,
and the two Lorentzian-based approximations agree.

At the second window, the analytical calculation fits the numerical solution for all chosen Doppler widths in the figure.
Eliminating~$I_2$,~(\ref{eq:I2}) leads to an equation for group velocity at the center of the second window
being similar equation to Eq.~(\ref{eq:DLF}) but with~$\Omega_\text{c}$ replaced by~$\Omega_\text{s}$ and~$\gamma_2$ replaced by~$\gamma_3$.
Similar dependence on Doppler width is shown in the inset of Fig.~\ref{fig:Dtemp}.

To achieve matched group velocity for the probe pulse propagating through the first and through the second window,
a non-zero nonlinearity is required.
The nonlinearity~$I_2$ is zero only if the condition $\rho_{44}=\rho_{33}=0$ is met.
This case for nonlinearity is depicted in the inset of Fig.~\ref{fig:Dtemp}. By fixing $\rho_{44}=\rho_{33}=0$ we have the unwanted additional effect of violating condition~(\ref{eq:hom2}) and thereby destroying the second window.

The intercept point between the two curves shown in Fig.~\ref{fig:Dtemp} reveals the operating temperature for group-velocity matching.
At temperatures exceeding the matched group-velocity case,
the group velocity in the first window is lower than the group velocity for the second window
and vice versa for temperatures lower than the condition for matched group velocity.

In summary, we demonstrate three important points in this subsection.
First, the Lorentzian approximation is a useful and valid approximation for studying the dispersion response of the probe field
as long as the conditions~(\ref{eq:inhom1}) and~(\ref{eq:inhom2}) for $I_2\equiv 0$
[Eq.~(\ref{eq:I2})] hold.
Second, the second term of Eq.~(\ref{eq:susplorent}) modifies the optical dispersion at the second window, which leads to a capacity for group velocity control through manipulating the temperature.
Finally, due to nonlinearity,
a signal-field intensity less than the coupling-field intensity does not necessarily imply
that the probe field has lower group velocity at the second EIT window than at the first window.

\subsection{Group-velocity reduction}
\label{sec:group}
In the previous Sec. \ref{subsec:width} and \ref{sec:Gvel} we have studied the behavior of the width and the group velocity for both EIT windows of the probe field in Doppler-broadening media. We have shown that a high-intensity coupling field is required to overcome inhomogeneous broadening, which represents an obstacle for group-velocity reduction. 
The width of the second EIT window is independent of temperature,
which means that the enhanced group-velocity reduction is superior to the case that would hold
if the width did depend on temperature as temperature dependence could only worsen this effect.

In this section, we derive two expressions that relate the signal-field Rabi frequency~$\Omega_\text{s}$
to the coupling-field Rabi~$\Omega_\text{c}$ and Doppler width~$W_\text{L}$.
Satisfying the first expression guarantees that the probe field has the same group velocity in each transparency window.
Satisfying the second expression guarantees the same HWHM for the two EIT windows.

The relation between~$\Omega_\text{s}$ and~$\Omega_\text{c}$ can be satisfied for a wide
range of temperatures bounded above and below by the requirements for the analytical approximations to be valid according to Eqs.~(\ref{eq:ValAppro}) and~(\ref{eq:inhom1}).
We then use these two expressions to divide the signal-field intensity to three regimes:
a low-strength regime where the group velocity and EIT width are lower than the first window,
a high-strength regime where both group velocity and width of EIT window are greater than for the first window,
and a middle regime where the group velocity is lower and the width is higher than for the first window.

\begin{figure} 
\includegraphics[width=1\columnwidth]{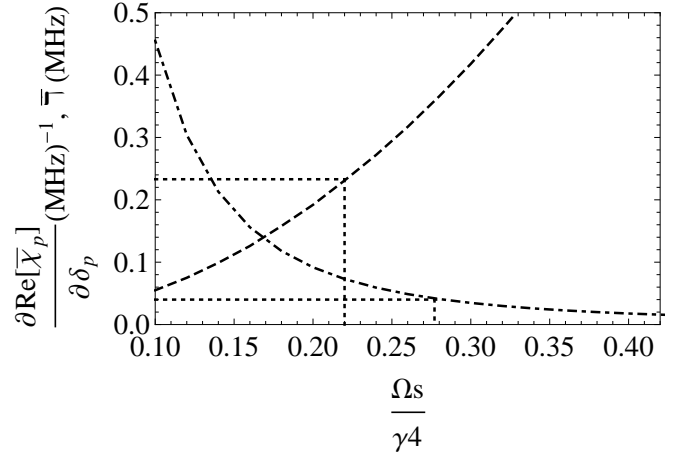}
\caption{%
	Plot of the partial derivative of dispersion (dotted-dashed line) and HWHM (dashed line) for the second EIT window
	and HWHM (upper horizontal dotted line line)
	and partial derivative of dispersion (lower horizontal dotted line line) for the first EIT window
	vs normalized signal-field Rabi frequency
	with $\Omega_\text{c}=\gamma_4$, $W_\text{L}=700$ MHz, 
 $\gamma_4=18$ MHz, $\gamma_3=10$ kHz, and $\gamma_2=40$ kHz.%
	}
\label{fig:Vreduction}
\end{figure}
In Fig.~\ref{fig:Vreduction},
we plot the HWHM and partial derivative of dispersion for the second EIT window
using Eqs.~(\ref{eq:width2}) and~(\ref{eq:DLS2}),
respectively.
We also plot the HWHM and partial derivative of dispersion for the first EIT window.
Intercepts between lines show which signal-field Rabi frequencies yield matched
HWHM or group-velocity conditions.
Matched HWHM occurs at $\Omega_{\text{s}_\text{l}}$
and matched group velocity occurs at $\Omega_{\text{s}_\text{h}}$
with $\Omega_{\text{s}_\text{l}}$ lower than $\Omega_{\text{s}_\text{h}}$,
and~l and~h refer to lower and higher values, respectively.
We can choose values of~$\Omega_\text{s}$ to control which of the two windows
has higher HWHM and group velocity.

We exploit our analytical expressions for the HWHM and the group velocity at the center of each window
to find the lower and higher boundary values of the signal field. 
Equating Eqs.~(\ref{eq:width1}) and~(\ref{eq:width2}) for real values
of~$\Omega_\text{c}$ and~$\Omega_\text{s}$ gives us the lower boundary value of~$\Omega_\text{s}$:
\begin{equation}
	\Omega_{\text{s}_\text{l}}=2^{\frac{3}{4}}\Omega_\text{c}\sqrt{\frac{\gamma_4}{W_\text{L}}}.
\label{eq:lowB}
\end{equation}
Similarly equating Eqs.~(\ref{eq:DLF}) and~(\ref{eq:DLS2})
gives us the higher boundary value of~$\Omega_\text{s}$:
\begin{equation}
  \Omega_{\text{s}_\text{h}}=\frac{2}{3}\sqrt{\frac{9}{2}\gamma_3W_\text{L}+\frac{\gamma_4\Omega_\text{c}^2}{3W_\text{L}}\left(19+\frac{2\gamma_4\Omega_\text{c}^2}{W_\text{L}^2\gamma_3}\right)}.
\label{eq:higB}
\end{equation}
Equations~(\ref{eq:lowB}) and~(\ref{eq:higB})
reveal which signal-field strength should be selected to achieve either matched width or matched group velocity, respectively.

At certain Doppler width, the boundary values of~$\Omega_\text{s}$ in Eqs.~(\ref{eq:lowB}) and~(\ref{eq:higB}) can be tuned by varying the coupling-field strength~$\Omega_\text{c}$.
Both the matched group velocity and the matched HWHM have lower value 
as~$\Omega_\text{c}$ is reduced.

In summary our four-level atom optical system can be operated at the second window
in three different regimes depending on the signal-field strength.
In the low-strength regime,
the second window has very low group velocity compared to the first window
but also has a lower EIT width. However, we can operate in this regime for lower group velocity as long as the width is resolvable experimentally.
Alternatively, in the high-strength regime,
the second window has a higher group velocity than for the first window.
which makes this high-strength regime less desirable for low group-velocity experiments.

\section{Conclusion}
\label{sec:summary}

We have achieved our objective of showing that the second DDEIT window 
has advantages over first window with respect to obtaining an enhanced reduction of group velocity.  
The presence of a nonlinear interaction between the probe and signal fields in optical susceptibility plays a crucial role in enabling temperature-controlled modification of the optical response.
At the second window, this term signifies the ability to reduce the narrowing of width and thereby yields increases of the group-velocity as the Doppler width increases.
The modified optical response due to nonlinear interaction permits observing the second window for low intense signal field and promises for more  group velocities' reduction in the second EIT window.

By identifying the signal-field boundary values~$\Omega_{\text{s}_\text{l}}$ and $\Omega_{\text{s}_\text{h}}$,
we are able to identify the regime of the signal-field strength values
that could result in slower group velocity  than for the first window.
The low-strength regime is the best for realizing low group velocity,
but the EIT window could be difficult to resolve.
The middle-strength regime is more robust in that the second EIT window is expected to be resolvable and the group velocity is expected to be low.
The high-strength regime is less interesting as the group velocity is relatively high.

Our approximate analytical calculation succeeds in describing the optical response
of the Doppler-broadened four-level optical system and helps in analyzing the system in the presence or absence of the nonlinear interaction. 
These analytical calculations also provide us with intuition of how the width or group velocities 
change in a Doppler-broadened system.  
Importantly, our analytical expression helps us to study the relation between the coupling and signal fields and to achieve matching of either widths or the group velocities of the two windows.
These conditions are not intuitively clear otherwise, and hence
would be difficult to discern using only numerical calculations.

\acknowledgments
We acknowledge financial support fromAlberta Innovates -Technology Futures (AITF) and Natural Sciences and Engineering Research Council of Canada (NSERC).
\appendix
\section{Atom-field Hamiltonian and equations of motion}
\label{sec:stationary}
The semiclassical Hamiltonian for the atom-field system depicted in Fig.~\ref{fig:tripod} ,
within the dipole and rotating-wave approximations,
is
\begin{equation}
	\hat{H}(t)=\hat{H}_0+\hat{H}_\text{dr}(t)
\end{equation}
with the free-atom Hamiltonian being
\begin{equation}
	\hat{H}_0
		=\hbar\sum_{\imath=1}^4\omega_\imath\hat{\sigma}_{\imath\imath},\;
	\hat{\sigma}_{\imath\jmath}:=\ket{\imath}\bra{\jmath},
\label{eq:Hfree}
\end{equation}
and the driving interaction Hamiltonian is
\begin{align}
	\hat{H}_\text{dr}(t)
		=&\frac{\hbar}{2}\Big(\Omega_\text{p}\text{e}^{\mathrm{i}\omega_\text{p}t}\hat{\sigma}_{14}\nonumber\\&
			+\Omega_\text{c}\text{e}^{\mathrm{i}\omega_\text{c}t}\hat{\sigma}_{24}
			+\Omega_\text{s}\text{e}^{\mathrm{i}\omega_\text{s}t}\hat{\sigma}_{34}
			+\text{H.c.}\Big),
\end{align}
with H.c.\ denoting the Hermitian conjugate.
In the interaction picture,
with respect to the free-atom Hamiltonian~(\ref{eq:Hfree}),
the atom-field system Hamiltonian has the form
\begin{align}
	\hat{V}(t)
		=&\frac{\hbar}{2}\Big(\Omega_\text{p}\text{e}^{-\mathrm{i}\delta_\text{p}t}\hat{\sigma}_{14}
					\nonumber\\&
			+\Omega_\text{c}\text{e}^{-\mathrm{i}\delta_\text{c}t}\hat{\sigma}_{24}
			+\Omega_\text{s}\text{e}^{-\mathrm{i}\delta_\text{s}t}\hat{\sigma}_{34}
			+\text{H.c.}\Big).
\end{align}

The Hamiltonian involves oscillatory terms at different optical frequencies.
Therefore, our next step is to find a Hermitian operator to transform the interaction Hamiltonian to a rotating frame in order to eliminate the time dependence. 
Thus,
we construct the rotating-frame operator
\begin{equation}
	\hat{A}=3\delta_\text{p}\hat{\sigma}_{11}
		+(2\delta_\text{p}+\delta_\text{c})\sigma_{22}
		+(2\delta_\text{p}+\delta_\text{s})\hat{\sigma}_{33}+2\delta_\text{p}\hat{\sigma}_{44}
\end{equation}
and eliminate time dependence by the following rotating-frame transformation
\begin{align}
	\hat{H}'(t)
		=\text{e}^{\text{i}\hat{A}t/\hbar}\hat{V}(t)\text{e}^{-\text{i}\hat{A}t/\hbar}-\hat{A}.
\end{align}
The resultant Hamiltonian is given by~\cite{Hessa2013}
\begin{align}
			\hat{H}'=&\hat{H}_0'
			+\frac{\hbar}{2}\Big(\Omega_\text{p}\hat{\sigma}_{14}
			+\Omega_\text{c}\hat{\sigma}_{24}+\Omega_\text{s}\hat{\sigma}_{34}+\text{H.c.}\Big)
\label{eq:resultantH}
\end{align}
for
\begin{equation}
	\hat{H}_0'
		:=\delta_\text{pc}\hat{\sigma}_{22}+\delta_\text{ps}\hat{\sigma}_{33}
			+\delta_\text{p}\hat{\sigma}_{44}
\label{eq:H0'}
\end{equation}
and~$\delta_\text{xy}$ given in Eq.~(\ref{eq:deltaxy}).

The resultant Lindblad master equation is~\cite{Hessa2013}
\begin{align}
	\dot{\rho}
		=&-\frac{\text{i}}{\hslash}[\rho,\hat{H}']+\sum_{\imath<\jmath}^4{\frac{\gamma_{\jmath\imath}}
			{2}(\sigma_{\imath\jmath}\rho\sigma_{\jmath\imath}-\sigma_{\jmath\jmath}\rho-\rho\sigma_{\jmath\jmath}})\nonumber\\
	&+\sum_{\jmath=2}^4{\frac{\gamma_{\phi j}}{2}(\sigma_{\jmath\jmath}\rho\sigma_{\jmath\jmath}-\sigma_{\jmath\jmath}\rho-\rho\sigma_{\jmath\jmath})}
\label{eq:masterequation}
\end{align}
for~$\rho$ the state in the rotating frame. The Lindblad master equation  includes both spontaneous emission and dephasing, where~$\gamma_{\jmath\imath}$ is the decay rate of state~$|\jmath\rangle\rightarrow|\imath\rangle$ and~$\gamma_{\phi\imath}$
is the dephasing of state~$|\imath\rangle$.

By substituting Eq.~(\ref{eq:resultantH}) into (\ref{eq:masterequation}),
we obtain 10 optical Bloch equations.
Six more optical Bloch equations are obtained from complex conjugates of the six off-diagonal 
density matrix expressions presented in the following. The 10 optical Bloch equations are
\begin{align}
\dot{\rho}_{12}(t)
		=&\left(-\frac{1}{2}\gamma_2+\text{i}\delta_\text{pc}\right)\rho_{12}(t)\nonumber\\
&-\frac{\text{i}}{2}\left[-\Omega^*_\text{c}\rho_{14}(t)
		+\Omega_\text{p}\rho_{24}(t)\right],\nonumber\\
	\dot{\rho}_{13}(t)
		=&\left(-\frac{1}{2}\gamma_3+\text{i}\delta_\text{ps}\right)\rho_{13}(t)\nonumber\\
		&-\frac{\text{i}}{2}\left[-\Omega^*_\text{s}\rho_{14}(t)
		+\Omega_\text{p}\rho_{43}(t)\right],
\nonumber\\
	\dot{\rho}_{14}(t)
		=&\left(-\frac{1}{2}\gamma_4+\text{i}\delta_\text{p}\right)\rho_{14}(t)\nonumber\\
		&+\frac{\text{i}}{2}\left[\Omega_\text{c}\rho_{12}+\Omega_\text{s}\rho_{13}
		+\Omega_\text{p}\left(\rho_{11}(t)-\rho_{44}(t)\right)\right],
\nonumber\\
	\dot{\rho}_{23}(t)
		=&\left(-\frac{1}{2}\Gamma_{32}
		-\text{i}\delta_\text{sc}\right)\rho_{23}(t)\nonumber\\
			&-\frac{\text{i}}{2}\left[\Omega_\text{c}\rho_{43}(t)-\Omega^*_\text{s}\rho_{24}(t)\right],
\nonumber\\
	\dot{\rho}_{24}(t)
		=&\left(-\frac{1}{2}\Gamma_{42}+\text{i}\delta_\text{c}\right)\rho_{24}(t)\nonumber\\
		&-\frac{\text{i}}{2}\left[-\Omega_\text{p}\rho_{21}(t)
		+\Omega_\text{c}\left(\rho_{44}(t)-\rho_{22}(t)\right)-\Omega_\text{s}\rho_{23}\right],
\nonumber\\
		\dot{\rho}_{43}(t)
			=&\left(-\frac{1}{2}\Gamma_{43}
			-\text{i}\delta_\text{s}\right)\rho_{43}(t)\nonumber\\
			&+\frac{\text{i}}{2}\big[-\Omega^*_\text{c}\rho_{23}(t)
			+\Omega^*_\text{s}(\rho_{44}(t)-\rho_{33}(t))\nonumber\\&
			-\Omega^*_\text{p}\rho_{13}(t)\big].
\label{eq:Droh14}
\end{align}
Now we present the equations of motion for the population density matrix elements:
\begin{align}
	\dot{\rho}_{11}(t)
		=&\gamma_{21}\rho_{22}(t)+\gamma_{31}\rho_{33}(t)+\gamma_{41}\rho_{44}(t)
						\nonumber\\&
		-\frac{\text{i}}{2}\Big[\Omega_\text{p}\rho_{41}(t)
						-\Omega^*_\text{p}\rho_{14}(t)\Big],\nonumber\\
	\dot{\rho}_{22}(t)
		=&-\gamma_{21}\rho_{22}(t)+\gamma_{32}\rho_{33}(t)
			+\gamma_{42}\rho_{44}(t)
						\nonumber\\&
			-\frac{\text{i}}{2}\left[-\Omega^*_\text{c}\rho_{24}(t)+\Omega_\text{c}\rho_{42}(t)\right],\nonumber\\
	\dot{\rho}_{33}(t)
	=&-\gamma_{31}\rho_{33}(t)-\gamma_{32}\rho_{33}(t)+\gamma_{43}\rho_{44}(t)
						\nonumber\\&
		-\frac{\text{i}}{2}\left[-\Omega^*_\text{s}\rho_{34}(t)+\Omega_\text{s}\rho_{43}(t)\right],\nonumber\\
	\dot{\rho}_{44}(t)
		=&-\gamma_{4}\rho_{44}(t)
			-\frac{\text{i}}{2}[\Omega_\text{c}\rho_{24}(t)-\Omega^*_\text{c}\rho_{42}(t)+\Omega_\text{s}\rho_{34}(t)\nonumber\\
	&-\Omega^*_\text{s}\rho_{43}(t)+\Omega_\text{p}\rho_{14}(t)
		-\Omega^*_\text{p}\rho_{41}(t)].
\label{eq:Droh44}
\end{align}
In summary, we reprise the master equation for a single tripod ($\pitchfork$)
atom driven by three detuned fields and obtain the requisite equations of motion
for the density-matrix elements to solve the dynamics.

\section{Dressed-state analysis and atomic Population}
\label{sec:Population}

Here, we analyze the population of the $\pitchfork$ atom
for various driving-field strengths and different detunings of the probe field.
In Sec.~\ref{subsec:population}, we investigate the atomic population behavior
using the dressed-state analysis and numerical calculations.
We obtain general expressions for atomic populations in steady state
for three cases of probe-field detuning.
In Sec.~\ref{subsec:atomicpopulation}, we introduce an analytical expression for the case we studied in this paper, corresponding to the signal-field strength being greater than the probe-field strength.
In Sec.~\ref{subsec:TemperaturePopulation},
we study the Doppler effect on the atomic population based on a numerical analysis.

\subsection{General discussion of atomic population}
\label{subsec:population}

Finding a general analytical expression for the atomic population using Eqs.~(\ref{eq:Droh14})
and~(\ref{eq:Droh44}) is not feasible due to the difficulties of decoupling the equations of motion of coherence from those of the population. Therefore, we analyze the dynamics of atomic population depending on the interpretation from dressed-state analysis and numerical calculation of the atomic population.

General expressions for the eigenstates of the Hamiltonian (\ref{eq:resultantH}) are complicated.
For simplicity, we choose to find the eigenstates for the three following tractable cases.

\subsubsection{$\delta_\text{pc}=0$}
\label{subsubsec:deltapc0}

The first case $\delta_\text{pc}=0$, or,
equivalently, $\delta_\text{p}=\delta_\text{c}$.
We allow~$\delta_\text{s}$ to assume any different value.
In this case one of the eigenvalues $\lambda_1=0$
corresponds to eigenstate
\begin{equation}
\label{eq:psiD}
	\ket{\psi_\text{D}}
		=-\frac{\Omega^*_\text{c}}{\sqrt{|\Omega_\text{c}|^2+|\Omega_\text{p}|^2}}\ket{1}+\frac{\Omega^*_\text{p}}{\sqrt{|\Omega_\text{c}|^2+|\Omega_\text{p}|^2}}\ket{2}.
\end{equation}
This eigenstate is a dark state as it does not contain a contribution from state~$|4\rangle$ and is not coupled to state~$|4\rangle$.
This is obvious from studying the total dipole moment~$\bm{d}_{4\text{D}}$
for a transition from state $\ket{\psi_\text{D}}$ to the bare state~$|4\rangle$.

If the magnitudes of the coupling field and probe field are appropriately balanced, 
the negative sign in the superposition of~$\ket{1}$ and~$\ket{2}$~[Eq.(\ref{eq:psiD})],
which forms the state $\ket{\psi_\text{D}}$,
causes the transition moment $\langle\psi_\text{D}|\bm{d}_{4\text{D}}|4\rangle$ to vanish.
Therefore,
if the atoms are formed in this state there is no possibility of excitation to~$|4\rangle$, hence no absorption.

For the case that the coupling field is much stronger than the probe field
($\Omega_\text{c}\gg\Omega_\text{p}$), state~$\ket{1}$
is almost equivalent to $\ket{\psi_\text{D}}$.
Thus, atoms decaying to state~$\ket{1}$ are trapped in this state and remain 
there throughout the interaction.
The atomic probability of being in state~$\ket{1}$
\begin{equation}
	P_1
		=\left|\langle1\ket{\psi_\text{D}}\right|^2
		=\frac{\left|\Omega_\text{c}\right|^2}{\left|\Omega_\text{c}\right|^2+|\Omega_\text{p}|^2}
\label{eq:pop1}
\end{equation}
and being in state~$\ket{2}$ is
\begin{equation}
	P_2
		=\left|\langle2\ket{\psi_\text{D}}\right|^2
		=\frac{|\Omega_\text{p}|^2}{\left|\Omega_\text{c}\right|^2+|\Omega_\text{p}|^2}.
\label{eq:pop2}
\end{equation}
\begin{figure}
\centering
\subfloat[\label{subfig-population1}]{%
      \includegraphics[width=0.9\columnwidth]{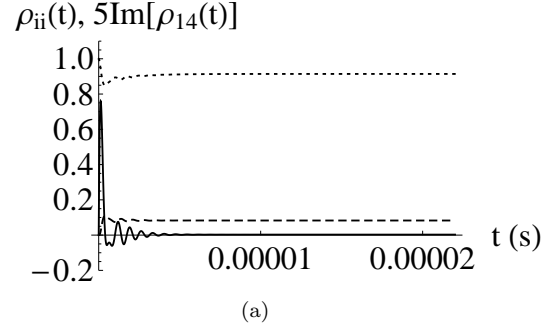}
}
\hfill
\subfloat[\label{subfig-population2}]{%
	\includegraphics[width=0.9\columnwidth]{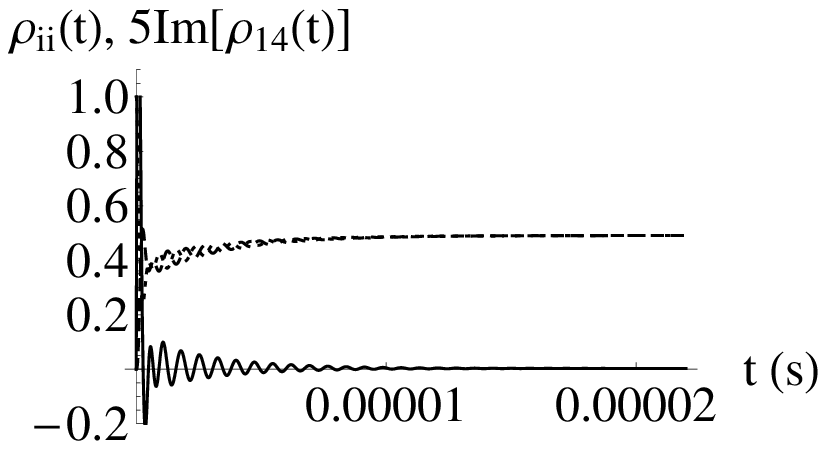}
       }
    \caption{%
    	Populations of state~$\ket{1}$ and~$\ket{2}$ represented by~$\rho_{11}$ (dotted line) and $\rho_{22}$ (dashed line) respectively, and absorption of the probe field represented by 5Im$[\rho_{14}]$ (solid line) 
	as a function of time~$t$
	evaluated numerically by solving the master equation.
	(a)~Coupling field is stronger
	than the probe field with $\Omega_\text{c}=\gamma_4$
	and $\Omega_\text{p}=0.3\gamma_4$.
	(b)~Coupling and probe fields have the same strength with $\Omega_\text{c}=\Omega_\text{p}=\gamma_4$. The system is initially prepared with $\rho^{(0)}_{11}=1$ and 
	$\rho^{(0)}_{22}=\rho^{(0)}_{33}=\rho^{(0)}_{44}=0$.
    The chosen parameters are $\gamma_4=18$ MHz, 
    $\gamma_3=10$ kHz, $\gamma_2=40$ kHz,
	$\Omega_\text{s}=0.3\gamma_4$, $\delta_\text{s}=0.5\Omega_\text{c}$, $\delta_\text{p}=\delta_\text{c}=0$.}
\label{fig:populationDeltapc}
\end{figure}

We numerically solve the master equation
and plot atomic populations in Fig.~\ref{fig:populationDeltapc}.
After a time of order of the radiative lifetime, the atoms should be trapped
in the dark state~$\ket{\psi_\text{D}}$,
which we verify by comparing the populations in Fig.~\ref{fig:populationDeltapc}
with the calculated dark-state populations.
Vanishing of the probe-field absorption Im$\rho_{14}$ supports the claim that the 
atom has decayed into a dark state.
Furthermore, state~$|4\rangle$ does not become populated.

For Fig.~\ref{fig:populationDeltapc}(a),~the dark state is equivalent to state~$\ket{1}$ whereas, for Fig.~\ref{fig:populationDeltapc}(b),
it is a superposition of states~$\ket{1}$ and~$\ket{2}$. The atoms are pumped into the state by combined action of coupling signal and probe field and spontaneous emission similar to optical pumping mechanism. At steady state, the distribution of atoms depends  mainly on the magnitude of the driving fields following the rule of Eqs.~(\ref{eq:pop1}) and~(\ref{eq:pop2}).
 
The other three eigenstates are
\begin{equation}
	\ket{\psi_\imath}=\frac{\Omega_\text{p} \ket{1}+\Omega_\text{c}\ket{2}+\frac{\lambda_\imath\Omega_\text{s}}{\lambda_\imath-\delta_\text{ps}}\ket{3}+ 2\lambda_\imath|4\rangle}{\sqrt{|\Omega_\text{p}|^2+\left|\Omega_\text{c}\right|^2+\frac{\left|\Omega_\text{s}\right|^2|\lambda_\imath|^2}{|\lambda_\imath-\delta_\text{ps}|^2}+4|\lambda_\imath|^2}}
\label{psiDeltapc}
\end{equation}
with eigenvalues $\lambda_\imath$ $(\imath\in2,3, 4)$
where each~$\{\lambda_\imath\}$ is a root of the eigenvalue equation 
\begin{align} 
	4\lambda^3-&4\lambda^2(\delta_\text{ps}+\delta_\text{p})+\lambda(4\delta_\text{ps}\delta_\text{p}-\left|\Omega_\text{c}\right|^2
-|\Omega_\text{p}|^2-\left|\Omega_\text{s}\right|^2)\nonumber\\
&+\delta_\text{ps}(\left|\Omega_\text{c}\right|^2+|\Omega_\text{p}|^2)=0.
\end{align}
In conclusion, detuning plays an important role in the distribution of atoms;
thus when $\delta_\text{pc}=0$ and after a time of the same order of atom relaxation time atoms  are trapped to the dark state and their distribution in the bare state~$\ket{1}$  and~$\ket{2}$ depends on the magnitude of $\Omega_\text{c}$ and $\Omega_\text{p}$, although the atoms are not prepared in the dark state. 

\subsubsection{$\delta_\text{ps}= 0$}
\label{subsubsec:deltaps0}

Now, we study the case that the probe and signal fields are at the two-photon resonance
with a $\ket{1}\leftrightarrow\ket{3}$ transition;
i.e., $\delta_\text{ps}=0$.
We allow the coupling detuning~$\delta_\text{c}$ to assume any value.
In this case, the Hamiltonian~(\ref{eq:resultantH}) has an eigenvalue $\lambda'=0$ with eigenstate
\begin{equation}
\label{eq:psi'D}
	\ket{\psi'_\text{D}}
		=\frac{-\Omega^*_\text{s}}{\sqrt{\left|\Omega_\text{s}\right|^2+|\Omega_\text{p}|^2}} \ket{1}+\frac{\Omega^*_\text{p}}{\sqrt{\left|\Omega_\text{s}\right|^2+|\Omega_\text{p}|^2}}\ket{3}
\end{equation}
and eignvalues $\lambda'_\imath$ $(\imath\in1,2, 3)$ 
with eigenstates 
\begin{equation}
	\ket{\psi'_\imath}
		=\frac{\Omega_\text{p} \ket{1}
		+\frac{\lambda'_\imath\Omega_\text{c}}
			{\lambda'_\imath-\delta_\text{pc}}\ket{2}+\Omega_\text{s}\ket{3}+ 2\lambda'_\imath|4\rangle}{\sqrt{\frac{\left|\Omega_\text{c}\right|^2\lambda'^2_\imath}{(\lambda'_\imath-\delta_\text{pc})^2}+|\Omega_\text{p}|^2+\left|\Omega_\text{s}\right|^2+4\lambda'^2_\imath}}
\label{eq:brightstate2}
\end{equation}
where each~$\lambda'_\imath$ is a root of the eigenvalue equation
\begin{align} 
	4\lambda'^3-&4\lambda'^2(\delta_\text{pc}+\delta_\text{p})+\lambda'(4\delta_\text{pc}\delta_\text{p}-\left|\Omega_\text{c}\right|^2-|\Omega_\text{p}|^2-\left|\Omega_\text{s}\right|^2)\nonumber\\
		&+\delta_\text{pc}(\left|\Omega_\text{s}\right|^2+|\Omega_\text{p}|^2)=0.
\end{align}
The eigenstate~$\ket{\psi'_\text{D}}$ is also a dark state as it does not  contain a contribution from state~$|4\rangle$ and is not coupled to state~$|4\rangle$. 

\begin{figure}
\centering
\subfloat[\label{subfig-population1}]{%
      \includegraphics[width=0.9\columnwidth]{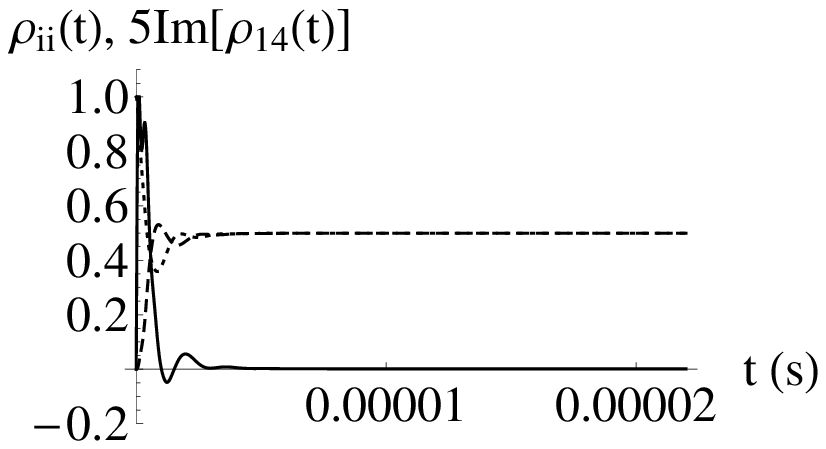}
}
\hfill
\subfloat[\label{subfig-population2}]{%
      \includegraphics[width=0.9\columnwidth]{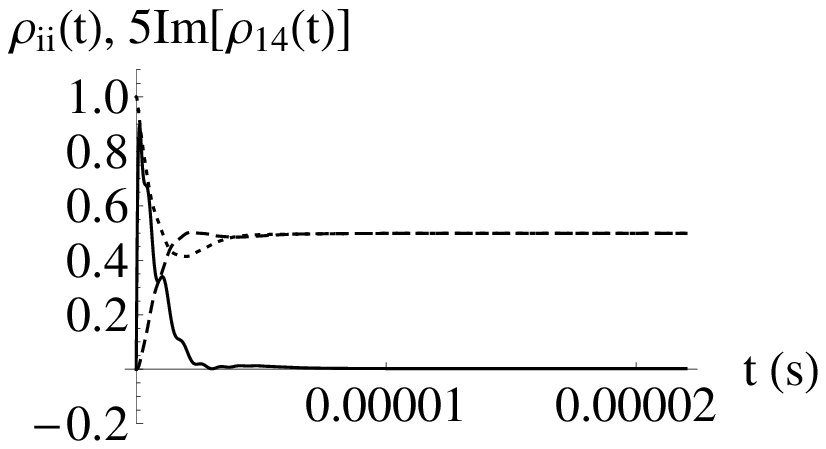}
}
\hfill
\subfloat[\label{subfig-population3}]{%
	\includegraphics[width=0.9\columnwidth]{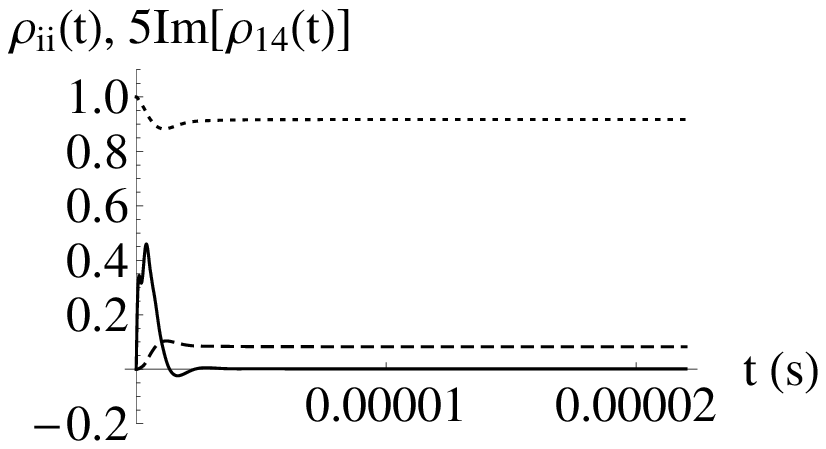}
}
	\caption{%
		Populations of level~$\ket{1}$ and~$\ket{3}$
		represented by~$\rho_{11}$ (dotted line) and~$\rho_{33}$ (dashed line), respectively,
		as a function of time~$t$
		evaluated numerically by solving the master equation.
		(a)~Signal- and probe-field strengths are of equal magnitude less than coupling field, with $\Omega_\text{s}=\Omega_\text{p}=0.5\gamma_4$ while $\Omega_\text{c}=\gamma_4$. (b)~Coupling-, signal- and probe-field strengths are of equal magnitude $\Omega_\text{c}=\Omega_\text{s}=\Omega_\text{p}=0.35\gamma_4$.
		(c)~Signal-field strength is stronger than the probe field, with $\Omega_\text{c}=\gamma_4$, $\Omega_\text{s}=0.5\gamma_4$, and $\Omega_\text{p}=0.15\gamma_4$.
    Other parameters are $\gamma_4=18$ MHz, 
    $\gamma_3=10$ kHz, $\gamma_2=40$ kHz,
	$\delta_\text{s}=\delta_\text{p}=0.5\Omega_\text{c}$, and $\delta_\text{c}=0$.
	Initial populations are $\rho^{(0)}_{11}=1$ and $\rho^{(0)}_{22}=\rho^{(0)}_{33}=\rho^{(0)}_{44}=0$. }
\label{fig:populationDeltaps}
\end{figure}

Atomic populations for states~$\ket{1}$ and~$\ket{3}$ 
are calculated numerically and shown in Fig.~\ref{fig:populationDeltaps}.
At steady state the atoms are trapped in the dark state~$\ket{\psi'_\text{D}}$ as long as the coupling field is  greater than or equal to the probe and to signal field.
We claim that the atom is trapped in the dark state because,
if instead the atom were in any one of the bright states of Eq.~(\ref{eq:brightstate2}),
the following phenomena would arise:
\begin{itemize}
	\item[(i)] We would expect to see some population in states~$\ket{2}$ and~$|4\rangle$
	whereas, in Figs.~\ref{fig:populationDeltaps}(a) and~(b),
	the populations of states~$\ket{1}$ and~$|3\rangle$ add almost to one,
	hence making the combined population of states~$\ket{2}$ and~$|4\rangle$ nearly zero.
	\item[(ii)] For the case that $\Omega_\text{c}\gg\Omega_\text{s}>\Omega_\text{p}$
	as shown in Fig.~\ref{fig:populationDeltaps}(c),
	if the system is in a bright state,
	then the population in state~$\ket{3}$ will exceed the population in state~$\ket{1}$,
	i.e., $\rho_{33}>\rho_{11}$,
	but the opposite is true:
	most of the population has been transferred to~$\ket{1}$.
	\item[(iii)] The absorption would not vanish for a bright state, but,
	in Figs. \ref{fig:populationDeltaps}(a)-\ref{fig:populationDeltaps}(c),
	absorption vanishes;
	hence, the atoms are trapped in the dark state~$\ket{\psi'_\text{D}}$. 
\end{itemize}
At steady state the populations in~$\ket{1}$ and~$\ket{3}$ are governed by the signal-
and probe-field Rabi frequencies according to
\begin{equation}
	P'_1
		=|\langle1|\psi'_\text{D}\rangle|^2
		=\frac{\left|\Omega_\text{s}\right|^2}{\left|\Omega_\text{s}\right|^2+|\Omega_\text{p}|^2}
\label{eq:pop1deltaps}
\end{equation}
and
 \begin{equation}
P'_3=|\langle3|\psi'_\text{D}\rangle|^2=\frac{|\Omega_\text{p}|^2}{\left|\Omega_\text{s}\right|^2+|\Omega_\text{p}|^2}.
\label{eq:pop2deltaps}
\end{equation}
respectively. 

\subsubsection{$\delta_\text{ps} =\delta_\text{pc}=0$}

The last case pertains to the three detunings are equal~\cite{Rebic2004},
($\delta_\text{p} = \delta_\text{c}=\delta_\text{s}$), which results in zero two-photon resonance.
Two of the eigenstates are degenerate eigenstates with eigenvalues
$\tilde{\lambda}_1=\tilde{\lambda}_2=0$:
\begin{align}
\label{eq:twodarkstates}
	\ket{\tilde{\psi}_\text{D1}}
		=&\frac{-\Omega^*_\text{s}}{\sqrt{\left|\Omega_\text{s}\right|^2
			+|\Omega_\text{p}|^2}}\ket{1}+\frac{\Omega^*_\text{p}}{\sqrt{\left|\Omega_\text{s}\right|^2
			+|\Omega_\text{p}|^2}}\ket{3}\nonumber\\
	\ket{\tilde{\psi}_\text{D2}}
		=&\frac{\Omega_\text{c}\Omega_\text{p} \ket{1}-(\Omega^2_p+\Omega^2_s)\ket{2}
			+\Omega_\text{c}\Omega_\text{s} \ket{3}}{\sqrt{\left(|\Omega_\text{c}|^2+|\Omega_\text{p}|^2
				+\left|\Omega_\text{s}\right|^2\right)\left(\left|\Omega_\text{p}\right|^2+|\Omega_\text{s}|^2\right)}}.
\end{align}
These two states are dark as neither contains contribution from state~$|4\rangle$
nor involve transitions to state~$|4\rangle$.
However, the rest of the eigenstates retain a component of all the bare atomic states:
\begin{equation}
	\ket{\tilde{\psi}^{\pm}}
		=\frac{\Omega_\text{p} \ket{1}+\Omega_\text{c}\ket{2}
			+\Omega_\text{s}\ket{3}\pm 2\tilde{\lambda}^{\pm}|4\rangle}{\sqrt{|\Omega_\text{c}|^2+|\Omega_\text{p}|^2+|\Omega_\text{s}|^2+4(\tilde{\lambda}^{\pm})^2}}
\end{equation}
with
\begin{equation}
	\tilde{\lambda}^\pm
		=\frac{1}{2}
			\left(
				\delta_\text{p} \pm\sqrt{\delta_\text{p}^2+\Omega_\text{p}^2+\Omega_\text{c}^2+\Omega_\text{s}^2}
			\right)
\end{equation}
the corresponding eigenvalues.

\begin{figure}
\centering
\subfloat[\label{subfig-population1}]{%
\setbox1=\hbox{3.cm}{\includegraphics[height=4cm]{{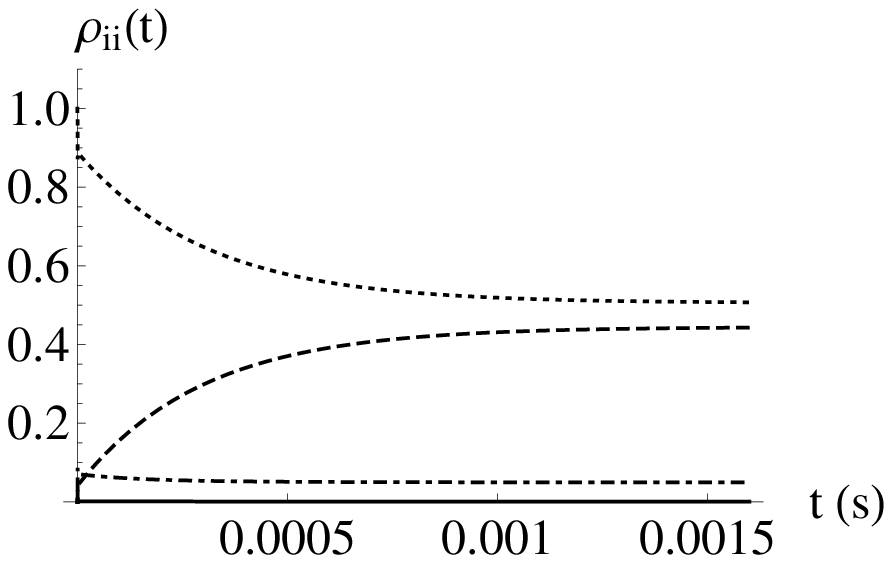}}}\llap{\raisebox{3.4cm}{\includegraphics[height=2.3cm]{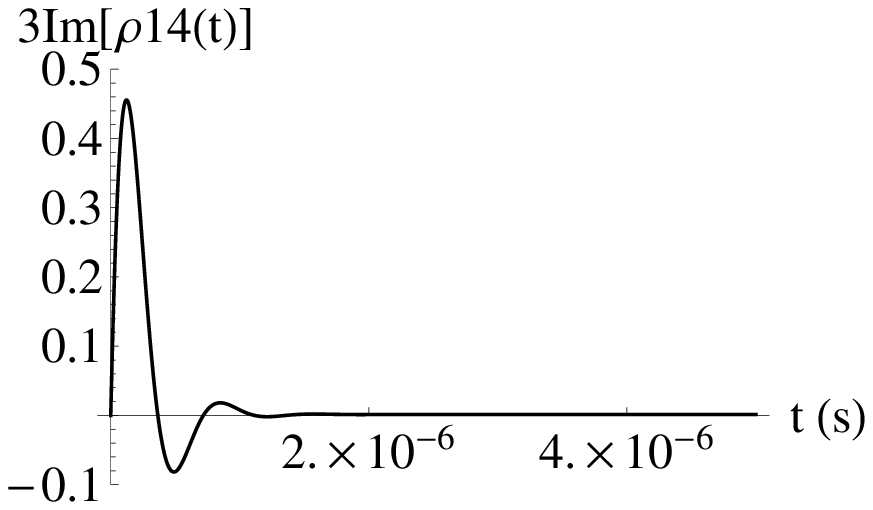}}}
}
\hfill
\subfloat[\label{subfig-population2}]{%
	\setbox1=\hbox{3.4cm}{\includegraphics[height=4cm]{{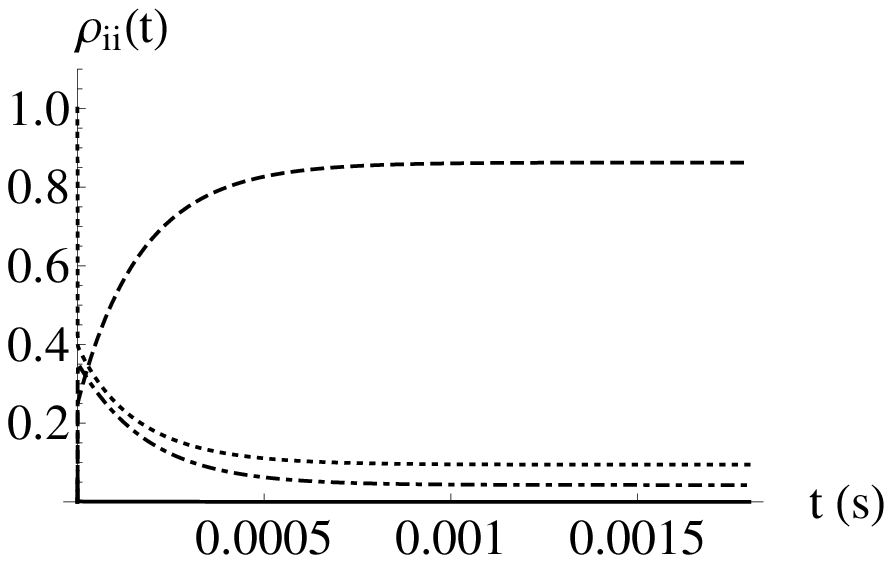}}}\llap{\raisebox{3.4cm}{\includegraphics[height=2.3cm]{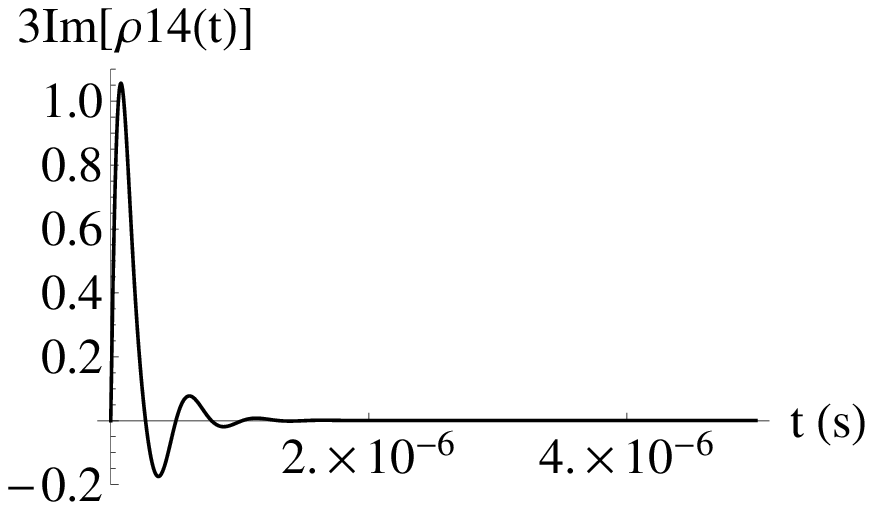}}}
}
\hfill
\subfloat[\label{subfig-population2}]{%
	\setbox1=\hbox{3.4cm}{\includegraphics[height=4cm]{{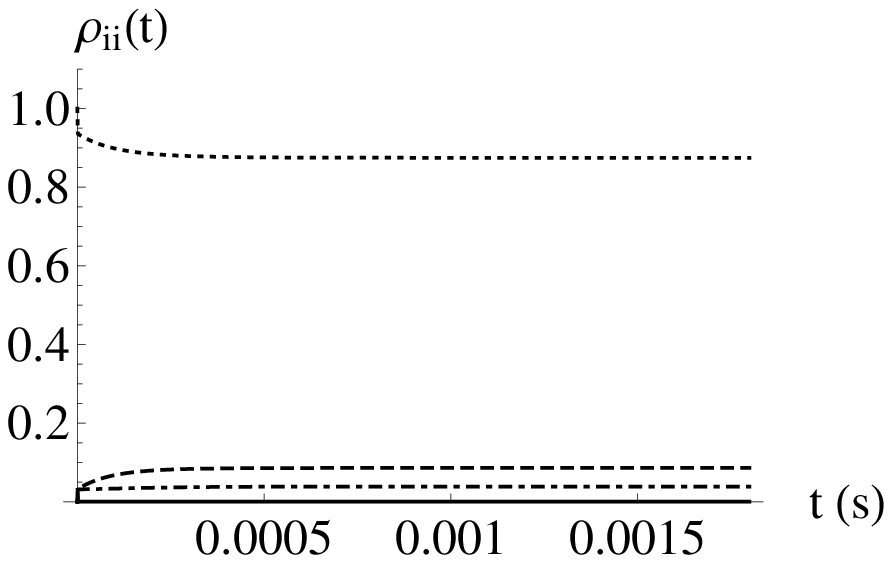}}}\llap{\raisebox{3.4cm}{\includegraphics[height=2.3cm]{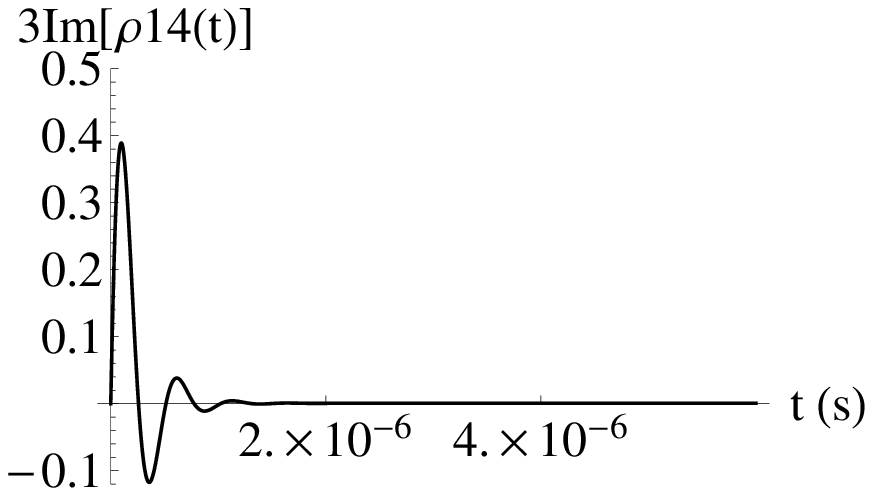}}}
}
\caption{%
	Populations of levels~$\ket{1}$,~$\ket{2}$ and~$\ket{3}$
	represented by~$\rho_{11}$ (dotted line), $\rho_{22}$ (dotted-dashed line), and~$\rho_{33}$ (dashed line), respectively,
	as a function of time~$t$
	evaluated numerically by solving the master equation.
	The conditions are
	(a)~$\Omega_\text{s}=\Omega_\text{p}=0.3\gamma_4$ and $\Omega_\text{c}=\gamma_4$,
	(b)~$\Omega_\text{p}, \Omega_\text{c}\gg\Omega_\text{s}$
	with $\Omega_\text{c}=\Omega_\text{p}=\gamma_4$ and $\Omega_\text{s}=0.3\gamma_4$, and
	(c)~$\Omega_\text{s}, \Omega_\text{c}\gg\Omega_\text{p}$ 
	with $\Omega_\text{c}=\Omega_\text{s}=1\gamma_4$ and $\Omega_\text{p}=0.3\gamma_4$.
	The initial population is $\rho^{(0)}_{11}=1$
	and $\rho^{(0)}_{22}=\rho^{(0)}_{33}=\rho^{(0)}_{44}=0$.
	Other parameters are $\gamma_4=18$ MHz,
	$\gamma_3=10$ kHz, $\gamma_2=40$ kHz, 
	$\delta_\text{p}=\delta_\text{s}=\delta_\text{c}=0$.
	Insets (a), (b), and (c) are the absorptions represented by Im$\rho_{14}$.%
	}
\label{fig:populationDeltapsc}
\end{figure}

The steady-state atomic populations for the case
\begin{equation}
	\delta_\text{p}=\delta_\text{s}=\delta_\text{c}=0
\end{equation} 
are shown Fig.~\ref{fig:populationDeltapsc}.
In all cases the atomic population is distributed between states~$\ket{1}$,~$\ket{2}$, and~$\ket{3}$
and excludes~$\ket{4}$.
This exclusion suggests that,
for all cases (a), (b), and (c),
atoms are trapped in the dark state~$\ket{\tilde{\psi}_\text{D2}}$,
but we see now that this guess could be true for case~(a) but not for cases~(b) and~(c).

In Fig.~\ref{fig:populationDeltapsc}(b),
we have $\Omega_\text{p}, \Omega_\text{c}\gg\Omega_\text{s}$ 
which means that, if the system is in dark state~$\ket{\tilde{\psi}_\text{D2}}$,
then the population in~$\ket{1}$ must be higher.
However, we see that the population of  state~$\ket{3}$ is in fact higher
and exhibits the opposite behavior to that shown in Fig.~\ref{fig:populationDeltapsc}(c).
Thus, the system corresponding to Figs.~\ref{fig:populationDeltapsc}(b)
and~\ref{fig:populationDeltapsc}(c) could be
trapped in~$\ket{\tilde{\psi}_\text{D1}}$. 
However, the few populations in state~$\ket{2}$ prevents us from making this conclusion as well.

From this argument, we conclude that the system is not in a pure dark state, but
relaxes into a mixture of the two dark states~(\ref{eq:twodarkstates})
which is also a dark state given by
\begin{equation}
\label{eq:mixeddarkstate}
	\tilde{\rho}_\text{D}
		=p_\text{D1}\ket{\tilde{\psi}_\text{D1}}\bra{\tilde{\psi}_\text{D1}}
			+p_\text{D2}\ket{\tilde{\psi}_\text{D2}}\bra{\tilde{\psi}_\text{D2}},
\end{equation}
where $p_\text{D1}$ is the probability of being in state $\ket{\tilde{\psi}_\text{D1}}$ and $p_\text{D2}$ is the probability of being in state $\ket{\tilde{\psi}_\text{D2}}$ such that
\begin{equation}
	p_\text{D1}+p_\text{D2}=1.
\label{eq:pd}
\end{equation}
The probability for state~$\ket{1}$ being populated is
\begin{align}
\label{eq:p1}
	\tilde{P}_1
		=&\left\langle 1\left|\tilde{\rho}_\text{D}\right|1\right\rangle\nonumber\\
		=&\frac{1}{\left|\Omega_\text{p}\right|^2+\left|\Omega_\text{s}\right|^2}\left(p_\text{D1}\left|\Omega_\text{s}\right|^2+\frac{p_\text{D2}\left|\Omega_\text{p}\right|^2\left|\Omega_\text{c}\right|^2}{\left|\Omega_\text{c}\right|^2+|\Omega_\text{p}|^2+|\Omega_\text{s}|^2}\right),
\end{align}
for populating state~$\ket{2}$ is
\begin{align}
\label{eq:p2}
	\tilde{P}_2
		=&\left\langle 2\left|\tilde{\rho}_\text{D}\right|2\right\rangle\nonumber\\
		=&\frac{p_\text{D2}\left|\Omega^2_\text{p}+\Omega^2_\text{s}\right|^2}{\left(\left|\Omega_\text{c}\right|^2+|\Omega_\text{p}|^2+|\Omega_\text{s}|^2\right)\left(\left|\Omega_\text{p}\right|^2+\left|\Omega_\text{s}\right|^2\right)},
\end{align}
and for populating state~$\ket{3}$ is
\begin{align}
\label{eq:p3}
	\tilde{P}_3
		=&\left\langle 3\left|\tilde{\rho}_\text{D}\right|3\right\rangle\nonumber\\
		=&\frac{1}{\left|\Omega_\text{p}\right|^2+\left|\Omega_\text{s}\right|^2}\left(p_\text{D1}\left|\Omega_\text{p}\right|^2+\frac{p_\text{D2}\left|\Omega_\text{s}\right|^2\left|\Omega_\text{c}\right|^2}{\left|\Omega_\text{c}\right|^2+|\Omega_\text{p}|^2+|\Omega_\text{s}|^2}\right).
\end{align}
The relation between $\tilde{P}_1$, $\tilde{P}_2$, and $\tilde{P}_3$ is
\begin{align}
	\tilde{P}_1+\tilde{P}_2+\tilde{P}_3=1
\end{align}
from Tr$\tilde{\rho}_\text{D}=1$.

We use numerical calculation of the population in state~$\ket{2}$
and Eq.~(\ref{eq:p2}) to determine the value of~$p_\text{D2}$. 
Once~$p_\text{D2}$ is known,
$p_\text{D1}$ is calculated from Eq.~(\ref{eq:pd}).
The agreement between the numerical values of~$\rho_{11}$ and~$\rho_{33}$
and the calculated values of~$\tilde{P}_1$ and~$\tilde{P}_3$ using Eqs.~(\ref{eq:p2}) and~(\ref{eq:p3}),
respectively, verify that the system is in a mixture of the two dark states~(\ref{eq:twodarkstates}).

We see that,
for certain two-photon detunings,
the system is eventually trapped in a dark state
even if the atom has not been prepared initially (at $t=0$)
in a dark state.
The atom is driven into the dark state
by a combined action of coupling, signal, and probe fields
and by spontaneous emission similar to that occurring through an optical pumping mechanism.

\begin{figure}
\centering
\subfloat[\label{subfig-stpopulation}]{%
      \includegraphics[width=.485\columnwidth]{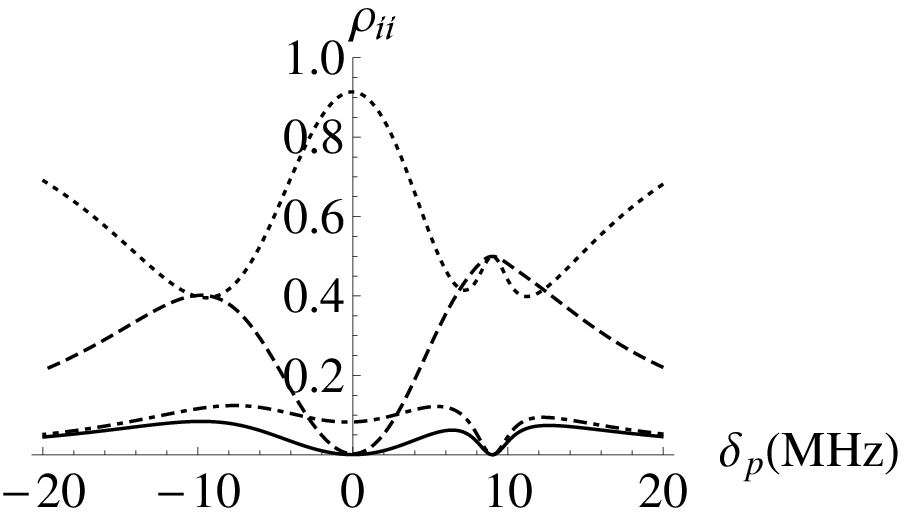}
}
\hfill
\subfloat[\label{subfig-stpopulation}]{%
      \includegraphics[width=.485\columnwidth]{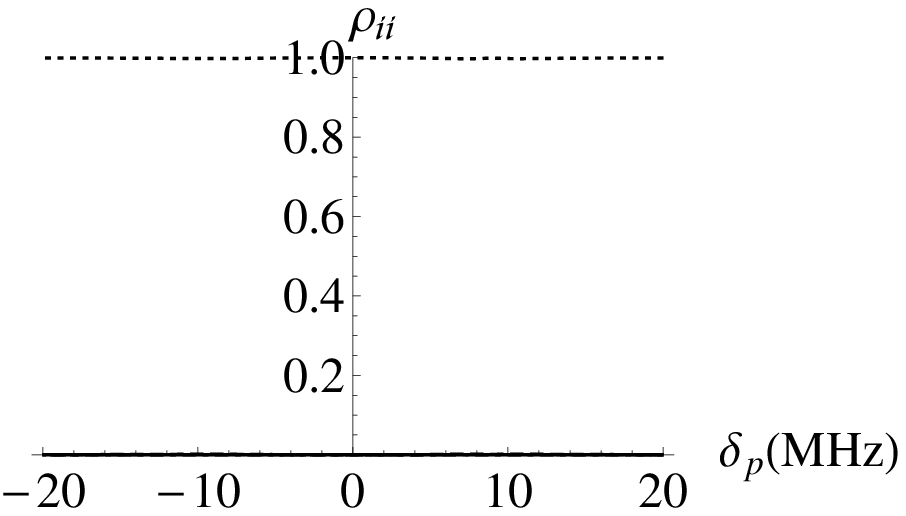}
}
\hfill
\subfloat[\label{subfig-DopPopulation}]{%
	\includegraphics[width=0.485\columnwidth]{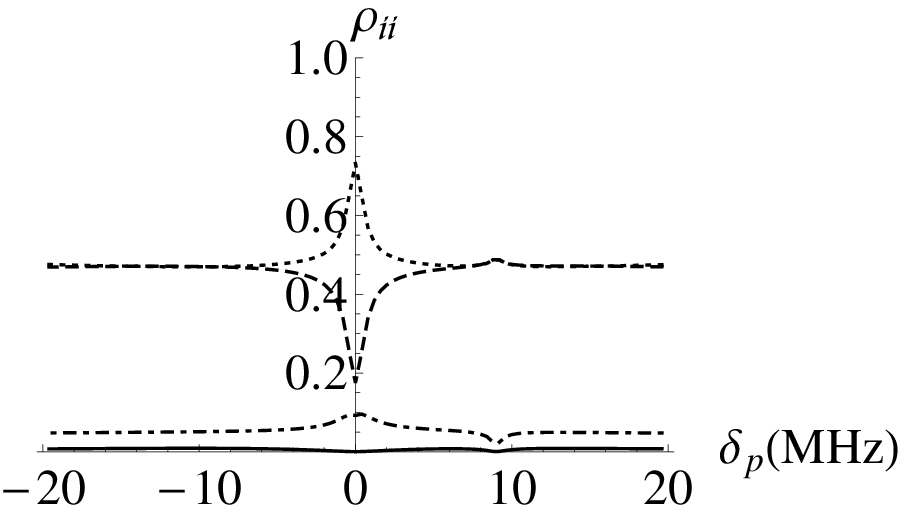}
       }
\hfill
\subfloat[\label{subfig-DopPopulation2}]{%
	\includegraphics[width=0.485\columnwidth]{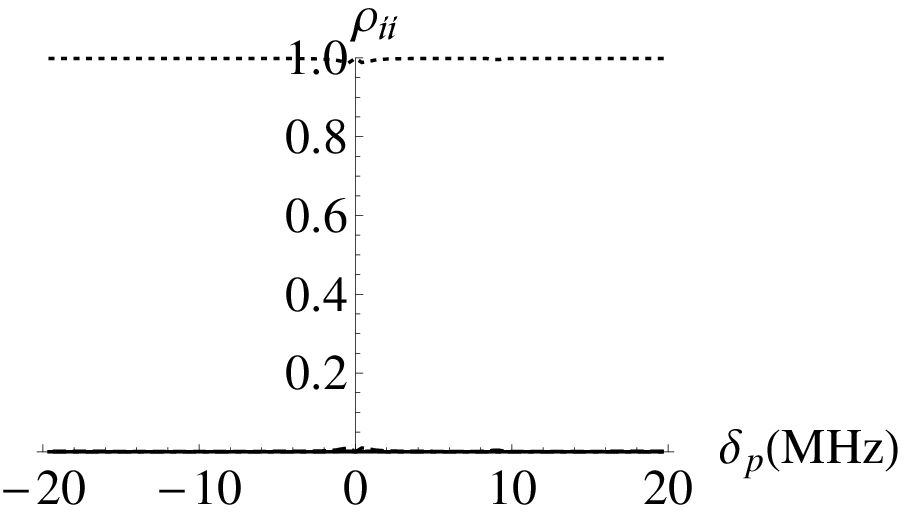}
       }
\caption{%
	Numerically evaluated steady-state populations 
	$\rho_{11}$ (dotted line), $\rho_{22}$ (dotted-dashed line), $\rho_{33}$ (dashed line), and $\rho_{44}$ (solid line)
	vs probe-field detuning~$\delta_{\text{p}}$
	evaluated (a),(b) at zero temperature and (c),(d) at 100 K.
	Parameter choices are
	(a) $\Omega_\text{s}=\Omega_\text{p}=0.3\gamma_4$. (b)~$\Omega_\text{s}\gg\Omega_\text{p}$,
	$\Omega_\text{s}=0.3\gamma_4$,  $\Omega_\text{p}=0.01\gamma_4$,
	(c) $\Omega_\text{s}=\Omega_\text{p}=0.3\gamma_4$,
	and (d) $\Omega_\text{s}\gg\Omega_\text{p}$, $\Omega_\text{s}=0.3\gamma_4$,  $\Omega_\text{p}=0.01\gamma_4$. 
	Other parameters are $\gamma_4=18$ MHz, 
    $\gamma_3=10$ kHz, $\gamma_2=40$ kHz, 
and $\Omega_\text{c}=\gamma_4$, $\delta_\text{s}=0.5\Omega_\text{c}$, $\delta_\text{c}=0$.}
\label{fig:populationfixDeltapsc}
\end{figure}

For stationary atoms (at low temperature),
the steady-state atomic population depends 
on probe-field detuning due to dark-state dependence on probe-field detuning.
Thus, changing the probe-field detuning modifies the steady-state population in each energy state,
as long as the probe field has comparable strength to the signal field
even if both are quite weak compared to the coupling field strength
as shown in Fig.~\ref{fig:populationfixDeltapsc}(a).
However, the dependence of the atomic population on probe-field detuning
decreases as the probe-field strength become weaker than the signal-field strength;
this feature is evident by comparing Fig.~\ref{fig:populationfixDeltapsc}(a)
with Fig.~\ref{fig:populationfixDeltapsc}(b).
Almost all of the population is evidently trapped in the dark state~$\ket{\psi_\text{D}}$ when $\delta_\text{pc}=0$ and in dark state~$\ket{\psi'_\text{D}}$ when $\delta_\text{ps}=0$.
which corresponds to state~$\ket{1}$ when $\Omega_\text{s}\gg\Omega_\text{p}$.

\subsection{Atomic population for probe-field strength weaker than signal-field strength}
\label{subsec:atomicpopulation}

In this section, we derive an analytical expression for atomic populations
for the case studied in our paper
corresponding to
$\Omega_\text{c}\gg\Omega_\text{s}\gg\Omega_\text{p}$.
The analytical expression can be found by solving Eqs.~(\ref{eq:Droh14}) and~(\ref{eq:Droh44})
restricted to the case that $\Omega_\text{p}\equiv 0$:
\begin{align}
\dot{\rho}_{23}(t)
		=&\left(-\frac{1}{2}\Gamma_{32}
		-\text{i}\delta_\text{sc}\right)\rho_{23}(t)-\frac{\text{i}}{2}\Omega_\text{c}\rho_{43}(t),
\nonumber\\
	\dot{\rho}_{24}(t)
		=&\left(-\frac{1}{2}\Gamma_{42}+\text{i}\delta_\text{c}\right)\rho_{24}(t)\nonumber\\
		&-\frac{\text{i}}{2}\left[
		\Omega_\text{c}\left(\rho_{44}(t)-\rho_{22}(t)\right)-\Omega_\text{s}\rho_{23}\right],
\nonumber\\
		\dot{\rho}_{43}(t)
			=&\left(-\frac{1}{2}\Gamma_{43}
			-\text{i}\delta_\text{s}\right)\rho_{43}(t)\nonumber\\
			&+\frac{\text{i}}{2}\left[-\Omega^*_\text{c}\rho_{23}(t)
			+\Omega^*_\text{s}(\rho_{44}(t)-\rho_{33}(t))
			\right],\nonumber\\
	\dot{\rho}_{11}(t)=&\gamma_{41}\rho_{44}(t)\nonumber\\
	\dot{\rho}_{22}(t)
		=&\gamma_{42}\rho_{44}(t)-\frac{\text{i}}{2}\left[-\Omega^*_\text{c}\rho_{24}(t)+\Omega_\text{c}\rho_{42}(t)\right],\nonumber\\
	\dot{\rho}_{33}(t)
	=&\gamma_{43}\rho_{44}(t)
		-\frac{\text{i}}{2}\left[-\Omega^*_\text{s}\rho_{34}(t)+\Omega_\text{s}\rho_{43}(t)\right],\nonumber\\
	1\equiv&\rho_{11}(t)+\rho_{22}(t)+\rho_{33}(t)+\rho_{44}(t).
\label{eq:denPopul1}
\end{align}
As we mentioned earlier the $\ket{1}\leftrightarrow\ket{2}$, $\ket{1}\leftrightarrow\ket{3}$,  and $\ket{2}\leftrightarrow\ket{3}$ transitions are dipole-forbidden.
Therefore, we restrict $\gamma_{21}=\gamma_{31}=\gamma_{32}=0$  in the above equation.

With initial population described by $\rho_{11}(0), \rho_{22}(0), \rho_{33}(0)$, and $\rho_{44}(0)$, the atomic populations for the four atomic bare states in the steady state are
\begin{align}
\label{eq:Popul1}
\rho_{11}=&\frac{\gamma_{41} Z  Y}{\gamma_{41} Z Y+\gamma_{43} (X-Y)},\nonumber\\
\rho_{22}=&\frac{\gamma_{41} Z \rho_{22}(0)+\left[Z (Y-\gamma_{42})-X\gamma_{43})(1-\rho_{11}(0)\right]}{\gamma_{41} Z Y+\gamma_{43} (X+Y)}\nonumber\\
&+\frac{X \gamma_{41} \rho_{33}(0)}{\gamma_{41} Z Y+\gamma_{43} (X-Y)},\nonumber\\
\rho_{33}=&\frac{Y \left[(Z +\gamma_{43 })(1-\rho_{11}(0))+\gamma_{41 }\rho_{33}(0)\right]}{\gamma_{41} Z Y+\gamma_{43} (X-Y)}\nonumber,\\
\rho_{44}=&\frac{Z(1-\rho_{11}(0)) Y}{\gamma_{41} Z Y+\gamma_{43} (X-Y)},
\end{align}
with
\begin{align}
\label{eq:contPop}
	X=&\frac{\left|\Omega_\text{s}\right|^2\left|\Omega_\text{c}\right|^2}{\left(\Gamma^2_{42}+4\delta^2_\text{c}\right)\left(\Gamma^2_{32}+4\delta^2_\text{sc}\right)}
					\nonumber\\&
	\left[\frac{\left(\Gamma_{43}+\frac{\left|\Omega_\text{c}\right|^2\Gamma_{32}}{\Gamma^2_{32}+4\delta^2_\text{sc}}\right)\left(4\delta_\text{s}\delta_\text{sc}+\Gamma_{32}\Gamma_{42}\right)}{\left(\Gamma_{43}+\frac{\left|\Omega_\text{c}\right|^2\Gamma_{32}}{\Gamma^2_{32}+4\delta^2_\text{sc}}\right)^2+4\left(\frac{\left|\Omega_\text{c}\right|^2\delta_\text{sc}}{\Gamma^2_{32}+4\delta^2_\text{sc}}-\delta_\text{s}\right)^2}\right.
\nonumber\\&
	\left.+\frac{\left(\frac{\left|\Omega_\text{c}\right|^2\delta_\text{sc}}{\Gamma^2_{32}+4\delta^2_\text{sc}}-\delta_\text{s}\right)\left(4\Gamma_{32}\delta_\text{c}-\Gamma_{42}\delta_\text{sc}\right)}{\left(\Gamma_{43}+\frac{\left|\Omega_\text{c}\right|^2\Gamma_{32}}{\Gamma^2_{32}+4\delta^2_\text{sc}}\right)^2+4\left(\frac{\left|\Omega_\text{c}\right|^2\delta_\text{sc}}{\Gamma^2_{32}+4\delta^2_\text{sc}}-\delta_\text{s}\right)^2}\right],
\nonumber\\
	Y=&\frac{\left|\Omega_\text{c}\right|^2\Gamma_{42}}{\Gamma^2_{42}+4\delta^2_\text{c}},
\end{align}
and
\begin{equation}
	Z=\frac{\left|\Omega_\text{s}\right|^2\left(\Gamma_{43}+\frac{\left|\Omega_\text{c}\right|^2\Gamma_{32}}{\Gamma^2_{32}+4\delta^2_\text{sc}}\right)}{\left(\Gamma_{43}+\frac{\left|\Omega_\text{c}\right|^2\Gamma_{32}}{\Gamma^2_{32}+4\delta^2_\text{sc}}\right)^2+4\left(\frac{\left|\Omega_\text{c}\right|^2\delta_\text{sc}}{\Gamma^2_{32}+4\delta^2_\text{sc}}-\delta_\text{s}\right)^2}.
\end{equation}
Equations~(\ref{eq:Popul1}) tell us that,
for all probe-field detunings,
almost all the atomic population is in state~$\ket{1}$
with almost no population in state~$\ket{3}$.
This lack of population in~$\ket{3}$
eliminates the effect of the nonlinear signal-probe interaction described by the second term of Eq.~(\ref{eq:rho143}).
As we require population in~$\ket{3}$,
we introduce an always-on incoherent pump at rate~$r$ to maintain population in~$\ket{3}$.

The equations of motion of the density matrix elements with the incoherent pumping are similar to those without incoherent pumping Eqs.~(\ref{eq:Droh14}) and~(\ref{eq:Droh44}),
differing only in the replacement
\begin{align}
\label{eq:repdecay}
	\gamma_4\rightarrow&\gamma_4+2r,\;
	\gamma_3\rightarrow\gamma_3+r,\nonumber\\
	\gamma_2\rightarrow&\gamma_2+r,\;
	\Gamma_{34}\rightarrow\Gamma_{34}+r.
\end{align}
The atomic population equations in the presence of the incoherent pumping are modified as
\begin{align}
\rho_{11} = \frac{Z Y \left(\gamma_{41} + r\right)}{Z \left(-r\gamma_{42} + 4rY + \gamma_{41}Y\right)+ r\gamma_{43} \left(Y- X\right)},\nonumber\\
\rho_{22}=\frac{r\left(Z\left(Y - \gamma_{42}\right) - \gamma_{43 }X\right)}{Z \left(-r\gamma_{42} + 4rY + \gamma_{41}Y\right)+ r\gamma_{43} \left(Y - X\right)},\nonumber\\
\rho_{33} =\frac{rY \left(Z + \gamma_{43}\right)}{Z \left(-r\gamma_{42} + 4rY + \gamma_{41}Y\right)+ r\gamma_{43} \left(Y - X\right)},\nonumber\\
\rho_{44} =\frac{rZY}{Z \left(-r\gamma_{42} + 4rY + \gamma_{41}Y\right)+ r\gamma_{43} \left(Y - X\right)}.
\end{align}
with~$X$,~$Y$, and~$Z$ defined in Eq.~(\ref{eq:contPop}) with replacement~(\ref{eq:repdecay}). 

\begin{figure}
\centering
\subfloat[\label{subfig-stPumpPopulation}]{%
      \includegraphics[width=.485\columnwidth]{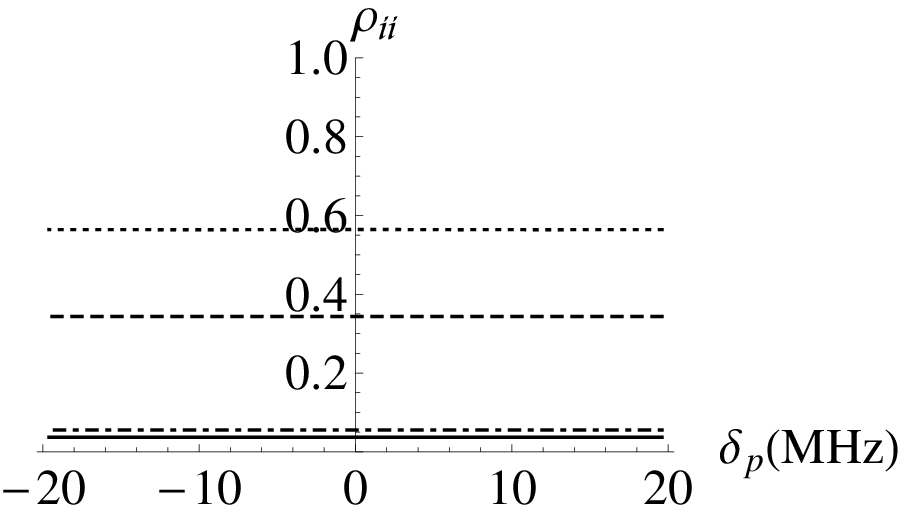}
}
\hfill
\subfloat[\label{subfig-stTPumpPopulation}]{%
\includegraphics[width=.485\columnwidth]{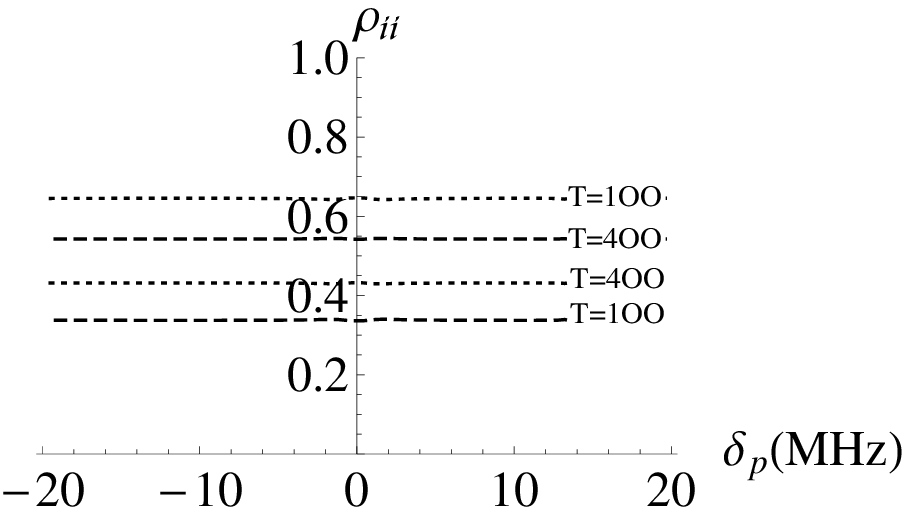}
}
\caption{%
	Steady-state populations
	$\rho_{11}$ (dotted line), $\rho_{22}$ (dot-dash),
	$\rho_{33}$ (dashed line), and $\rho_{44}$ (solid line) 
	vs probe-field detuning~$\delta_{\text{p}}$
	for $\Omega_\text{s}\gg\Omega_\text{p}$ in the presence of incoherent pumping with constant rate~$r$
    evaluated numerically by solving the master equation.
    The solutions correspond to
    (a)~low temperature for $r=1$MHz
    and (b)~for two temperatures 100 and 400 K with $r=0.01$MHz.
    Other parameters are $\Omega_\text{s}=0.3\gamma_4$,  $\Omega_\text{p}=0.01\gamma_4$,
    $\gamma_{41}=\gamma_{42}=6$MHz, $\gamma_{43}=12$MHz
    $\gamma_3=10$ kHz, $\gamma_2=40$ kHz, 
and $\Omega_\text{c}=\gamma_4$, $\delta_\text{s}=13.5$ MHz, $\delta_\text{c}=0$.}
\label{fig:stpopulationPump}
\end{figure}

The modified atomic population in the presence of incoherent pumping for the case
$\Omega_\text{s}\gg\Omega_\text{p}$ is shown in Fig.~\ref{fig:stpopulationPump}(a).
The existence of incoherent pumping makes the population constant for all probe detunings.
The value of the pumping rate~$r$ controls the population in each state. We use $r=1$ MHz to populate 
states~$\ket{1}$ and~$\ket{3}$ with $\rho_{11}=\rho_{33}=0.44$.
At high temperature,
when the Doppler effect plays a critical role in repopulating the states,
$\ket{3}$ can be repopulated to a value of half
using pump rate~$r$ as low as 1kHz
as shown in Fig.~\ref{fig:stpopulationPump}(b).

In this section,
we have presented an incoherent pumping procedure that maintains equal population
between~$\ket{1}$ and~$\ket{3}$ for a weak probe field.
In the main body of the paper,
we did not treat pumping; 
instead, we assumed  equal population between~$\ket{1}$ and~$\ket{3}$.
As we use an incoherent pump,
we should be concerned that coherence is affected,
but we see here that dephasing due to incoherent pumping is negligible for reasonable parameters.
Specifically, the extra dephasing of system from incoherent pumping
is of the same order as~$\gamma_{\phi3}$ for Doppler-broadened media~\cite{MCL08}.

\subsection{Temperature dependence of atomic population  }
\label{subsec:TemperaturePopulation}
Increasing the temperature adds two more phenomena to the atom-field system
which we incorporated into an extended quantum master equation.
These two phenomena are thermal dissipation and Doppler broadening.
Our examination of thermal dissipation shows that its effect is too weak
to influence substantially either the coherence or the population.
However, the second phenomenon of Doppler broadening modifies the coherence,
as presented from Sec.~\ref{sec:Doppler} onward for equal populations in
levels~$\ket{1}$ and~$\ket{3}$,
is discussed in this subsection in the absence of this equal-population restriction.

Finding an analytical expression for population in Doppler-broadening medium is difficult.
Therefore, we perform numerical studies of atomic populations for various temperatures.
Now, we proceed to analyze the connection between atomic population and coherence.

At high temperature,
Doppler broadening reduces coherence~\cite{Ackerhalt79,Choe1995}
and specifically directly reduces the coherences~$\rho_{14}$ and~$\rho_{34}$ 
that are established by the weak fields.
Consequently the populations of the states~$\ket{1}$ and~$\ket{3}$
change according to the solution of Eq.~(\ref{eq:Droh44}).
We see from the approximate analytical expression for optical susceptibility~(\ref{eq:susplorent})
that increasing Doppler width~$W_\text{L}$ in Eq.~(\ref{eq:susplorent})
is responsible for reducing coherence.

\begin{figure}
\centering
\subfloat[\label{subfig-PopulationHigDep}]{%
\includegraphics[width=.485\columnwidth]{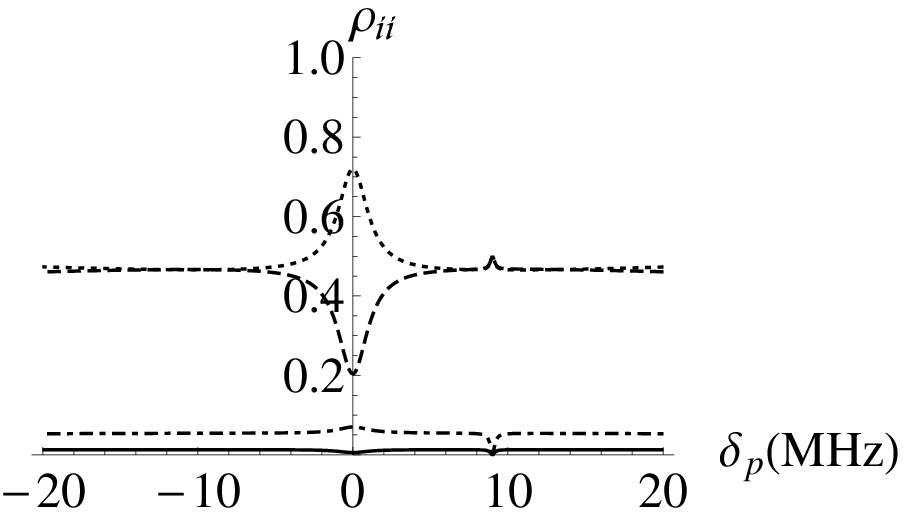}
} 
\hfill
\subfloat[\label{subfig-PopulationHigDep2}]{%
\includegraphics[width=.485\columnwidth]{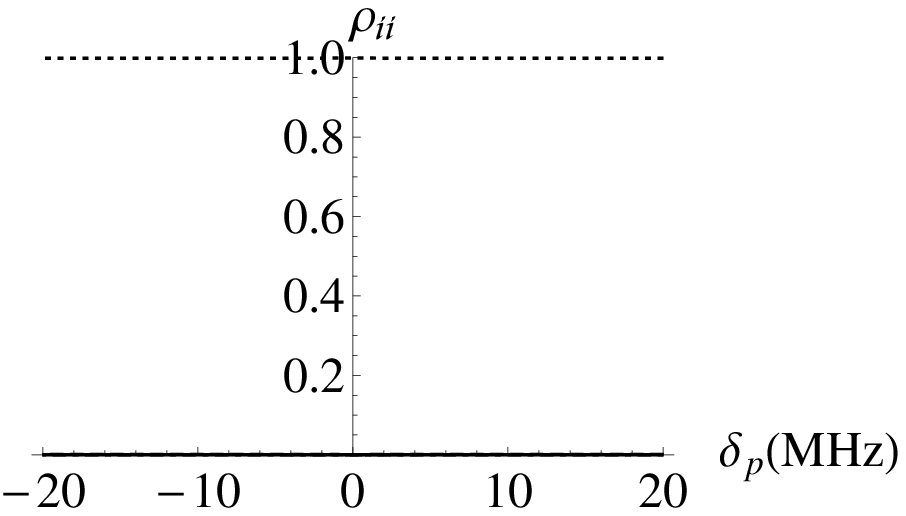}}
\caption{%
	Numerically evaluated steady-state populations 
	$\rho_{11}$ (dotted line), $\rho_{22}$ (dot-dash),	$\rho_{33}$ (dashed line), and $\rho_{44}$ (solid line)
	vs probe-field detuning~$\delta_{\text{p}}$
	at zero temperature and incorporating dephasing $\gamma_{\phi41}=\gamma_{\phi43}=$150MHz  and $\gamma_{\phi2}=$0.8MHz.
	The plots show
	(a)~$\Omega_\text{s}=\Omega_\text{p}=0.3\gamma_4$
	and (b)~$\Omega_\text{s}\gg\Omega_\text{p}$, $\Omega_\text{s}=0.3\gamma_4$,  $\Omega_\text{p}=0.01\gamma_4$.
	 Other parameters are $\gamma_4=18$ MHz, $\gamma_3=10$ kHz,
and $\Omega_\text{c}=\gamma_4$, $\delta_\text{s}=0.5\Omega_\text{c}$, $\delta_\text{c}=0$.%
	}
\label{fig:populationHigDeph}
\end{figure}

Reduction of coherence and its consequent effects
due to Doppler broadening and to increasing Doppler width
are similar to the effects due to adding extra dephasing~$\gamma_{\phi41}$
between~$\ket{1}$ and~$\ket{4}$ and $\gamma_{\phi43}$ 
between~$\ket{3}$ and~$\ket{4}$
plus increasing the dephasing~$\gamma_{\phi 2}$ between~$\ket{1}$ and~$\ket{2}$.
In Fig.~\ref{fig:populationHigDeph}, we present numerically evaluated atomic populations
for different probe field detunings~$\delta_{\text{p}}$ at zero temperature
accompanied by additional dephasing quantified by~$\gamma_{\phi41}$ and by~$\gamma_{\phi43}$.
We choose the dephasing $\gamma_{\phi41}$  and $\gamma_{\phi43}$ in Fig.~\ref{fig:populationHigDeph} to be of the same order of magnitude of $W_\text{L}$.
We observe that Figs.~\ref{fig:populationfixDeltapsc}(c), and \ref{fig:populationfixDeltapsc}(d) 
are similar to Figs.~\ref{fig:populationHigDeph}(a), and \ref{fig:populationHigDeph}(b).

\subsubsection{$\Omega_\text{s}=\Omega_\text{p}$}

At $\delta_\text{p}=\delta_\text{c}=0$, the population of atoms in state~$\ket{3}$ increases as temperature increases, as shown Fig.~\ref{fig:populationfixDeltapsc}(c). 
Increasing the temperature from 0 to 100 K decreases~$\rho_{11}$ from~1.0 to~0.7
while increasing~$\rho_{33}$ from~0.0 to~0.2.
This population changes is due to reduction of the coherence $\rho_{14}$.
The coherence $\rho_{34}$ does not have an influence at $\delta_\text{pc}=0$. 

The atom-field system is no longer in the pure dark state $\ket{\psi_\text{D}}$.
Increasing the temperature has the effect of displacing the system
from the dark state to different state, where the absorption of the probe field increases.
Thus, as the temperature of the system increases,
probe-field absorption becomes very high, thereby potentially preventing the first EIT window from being observed.

For $\delta_\text{p}=\delta_\text{s}$, the atomic
population remains the same at different temperatures;
i.e., $\rho_{11}=\rho_{33}=0.5$.
Therefore, the system remains trapped in the dark state $\ket{\psi'_\text{D}}$ [Eq. (\ref{eq:psi'D})].
The population at that detuning is less sensitive to Doppler broadening.
This insensitivity can be explained as resulting from
higher-order nonlinear interactions between the signal and probe fields
resulting from the coupling of~$\rho_{14}$ to~$\rho_{34}$
through the presence of the term $\rho_{13}$~\cite{Hessa2013}.
This coupling eliminates the effect of Doppler broadening and eliminates the dephasing by $W_\text{L}$.
Thus, at the second window, the dressed atom-field dark state is stable 
with respect to the Doppler effect,
and this stability enables observing the second window
even at higher temperature.

\subsubsection{$\Omega_\text{s}\gg\Omega_\text{p}$}
For $\Omega_\text{s}\gg\Omega_\text{p}$,
the term $\Omega_\text{p}\rho_{41}(t)-\Omega^*_\text{p}\rho_{14}(t)$ of Eq.~(\ref{eq:Droh44}) is neglected because its effect is very small.
Therefore, for this case,
the coherence $\rho_{14}$ has negligible effect on the atomic population of state~$\ket{1}$.
Consequently, the population of~$\ket{1}$ is not affected by Doppler broadening
at $\delta_{\text{p}}=\delta_{\text{c}}$.

Under the additional constraint that $\delta_\text{ps}=0$,
only $\rho_{34}$ affects the population by reducing $\rho_{33}$, and,
as~$\rho_{33}$ for the Doppler-free atomic-field system is almost zero,
reducing $\rho_{34}$ thus has no effect on the population of state~$\ket{3}$.
Hence, the population of each level~(\ref{eq:Droh44})
at zero temperature will be the same as for the population at any higher temperature.
At steady state, the atoms are all trapped in the state~$\ket{\psi_\text{D}}$
for $\delta_\text{pc}=0$ and to~$\ket{\psi'_\text{D}}$ for $\delta_\text{ps}=0$.

\section{Absorption maxima and minima}
\label{sec:minmax}
In this section, we show the calculations leading to closed-form expressions for 
the heights, or maxima, of the two absorption windows and also the nadirs, or minima,
of these absorption windows.
These expressions are derived first for the stationary atom and then generalized to the Doppler-broadened case.

\subsection{Stationary atom}
\label{subsec:stationaryatom}
Identifying the maximum height,~$h_{\text{max}_\imath}$ and minimum height,~$h_{\text{min}_\imath}$
of the $\imath^\text{th}$ window is subtle because the two Lorentzian transparency windows are 
cut asymmetrically into the overall Lorentzian absorption peak corresponding to zero-coupling field.
First, we consider the $\imath=1$ case.

The maximum is calculated by setting $\Omega_\text{c}=0=\Omega_\text{s}$ and evaluating
Eq.~(\ref{eq:suscep}) at $\delta_\text{p}=\delta_\text{c}=0$:
\begin{equation}
	h_{\text{max}_1}=\eta\frac{\rho_{11}-\rho_{44}}{\gamma_4}.
\end{equation}
The minimum $h_{\text{min}_1}$ is determined by setting $\Omega_\text{s}=0$ but $\Omega_\text{c}\neq0$ and evaluating Eq.~(\ref{eq:suscep}) at $\delta_\text{p}=\delta_\text{c}=0$. 
We obtain
\begin{equation} 
	h_{\text{min}_1}
		=\frac{\eta\left(\rho_{11}-\rho_{44}\right)\gamma_2}
			{\gamma_4\gamma_2+\left|\Omega_\text{c}\right|^2}
		\xrightarrow{\scriptscriptstyle \gamma_2\to 0} 0
\end{equation} 
with zero absorption attained for $\gamma_2=0$.
If $\gamma_2\neq 0$ but condition~(\ref{eq:hom1}) holds,
\begin{equation}
	h_{\text{min}_1}\to\frac{\eta\left(\rho_{11}-\rho_{44}\right)\gamma_2}{\left|\Omega_\text{c}\right|^2},
\end{equation} 
and minimum absorption is reached.

The maximum and minimum of the first transparency window
are used to calculate the half-maximum
\begin{equation}
	\varkappa_1
		:=\frac{h_{\text{max}_1}
				+h_{\text{min}_1}}{2}
		=\frac{\eta\left(\rho_{11}-\rho_{44}\right)\left(2\gamma_4
			\gamma_2+\left|\Omega_\text{c}\right|^2\right)}{2\gamma_4\left(\gamma_4\gamma_2+\left|\Omega_\text{c}\right|^2\right)}.
\end{equation} 
Applying condition~(\ref{eq:hom1}) yields
\begin{equation}
	\varkappa_1=\frac{\eta\left(\rho_{11}-\rho_{44}\right)}{2\gamma_4}.
\end{equation} 

For the $\imath=2$ case,
we have a Lorentzian transparency window cut into the absorption curve corresponding
to the two conditions $\Omega_\text{s}=0$ and $\Omega_\text{c}\neq 0$ holding.
For our case of DDEIT, we set $\Omega_\text{c}=2\delta_\text{s}$,
which establishes the second transparency window centered at $\delta_\text{p}=\delta_\text{s}$,
which is the point that the maximum peak height~$h_{\text{max}_2}$ occurs.
Therefore, $h_{\text{max}_2}$ can be determined
by calculating the $\Omega_\text{s}=0$ absorption curve value at~ $\delta_\text{p}=\delta_\text{s}$:
\begin{equation}
h_{\text{max}_2}=\eta\frac{\left(\rho_{11}-\rho_{44}\right)\left(\gamma_4+\frac{\Omega^2_\text{c}\gamma_2}{4\delta^2_\text{sc}}\right)}{\gamma^2_4+4\left(\frac{\Omega^2_\text{c}}{4\delta^2_\text{sc}}-\delta_\text{s}\right)^2}.
\end{equation}
Under the approximation that 
\begin{equation}
\label{eq:smallerthangamma4}
	\gamma_4\gg\frac{\left|\Omega_\text{c}\right|^2}{2\delta^2_\text{sc}}\frac{\gamma_2}{2},
\end{equation}
we obtain
\begin{equation}
	h_{\text{max}_2}=\eta\frac{\rho_{11}-\rho_{44}}{\gamma_4}.
\end{equation}
Thus,
$h_{\text{max}_1}\approx h_{\text{max}_2}$.

In order to calculate the minimum of the second transparency window,
we set $\Omega_\text{s}\neq0$ and $\Omega_\text{c}\neq0$
and evaluate Im$\chi^{(1)}_\text{p}$ from Eq.~(\ref{eq:suscep}) at $\delta_\text{p}=\delta_\text{s}$
to obtain the minimum
\begin{equation}
	h_{\text{min}_2}=\eta\left(\frac{\left(\rho_{11}-\rho_{44}\right)\gamma_3}{\gamma_4\gamma_3+\left|\Omega_\text{s}\right|^2}+\frac{\left(\rho_{44}-\rho_{33}\right)\left|\Omega_\text{s}\right|^2}{\Gamma_{43}\left(\gamma_4\gamma_3+\left|\Omega_\text{s}\right|^2\right)}\right).
\label{min2}
\end{equation}
The first term in the right-hand side
of Eq.~(\ref{min2}) represents the absorption minimum,
whereas the second term represents the maximum gain (negative absorption).

If we wish to reduce absorption,
decay from level~$\ket{3}$ must be minimized,
i.e., $\gamma_3\rightarrow 0$.
For the case $\gamma_3\neq0$, condition~(\ref{eq:hom2})
must be satisfied to minimize the absorption. 
As $\Gamma_{43}\approx\gamma_4$,
Eq.~(\ref{min2}) is simplified to
\begin{equation}
h_{\text{min}_2}=\eta\left(\frac{\left(\rho_{11}-\rho_{44}\right)\gamma_3}{\left|\Omega_\text{s}\right|^2}+\frac{\rho_{44}-\rho_{33}}{\gamma_4}\right).
\label{min2f}
\end{equation}
As $\rho_{44}=0$ is assumed,
gain exists only when
\begin{equation}
	\frac{\left|\rho_{44}-\rho_{33}\right|}{\gamma_4}\gg\frac{\left|\rho_{11}-\rho_{44}\right|\gamma_3}{\left|\Omega_\text{s}\right|^2}
\end{equation}
or, equivalently, if
\begin{equation}
	\frac{\left|\rho_{44}-\rho_{33}\right|}{\left|\rho_{11}-\rho_{44}\right|}\gg\frac{\gamma_3\gamma_4}{\left|\Omega_\text{s}\right|^2}.
\end{equation}

In our case
\begin{equation}
	\frac{\left|\rho_{44}-\rho_{33}\right|}{\left|\rho_{11}-\rho_{44}\right|}\approx1.
\end{equation}
Therefore, we also require condition~(\ref{eq:hom2}) in order to achieve gain.
The half-maximum is then
\begin{align}
	\varkappa_2
		=&\eta\left[\frac{\left(\rho_{11}-\rho_{44}\right)\left(2\gamma_4\gamma_3+\left|\Omega_\text{s}\right|^2\right)}{\gamma_4\left(\gamma_4\gamma_3+\left|\Omega_\text{s}\right|^2\right)}\right.
				\\ \nonumber&
		+\left.\frac{\left(\rho_{44}-\rho_{33}\right)\left|\Omega_\text{s}\right|^2}{\Gamma_{43}\left(\gamma_4\gamma_3+\left|\Omega_\text{s}\right|^2\right)}\right].
\end{align} 
For condition~(\ref{eq:hom2}) and~$\gamma_4\gg\gamma_3$,
\begin{equation}
	\varkappa_2=\eta\frac{\left(\rho_{11}-\rho_{33}\right)}{\gamma_4}.
\end{equation}
For $\rho_{11}\approx\rho_{33}$,
$\varkappa_2$ is located at zero absorption.

\subsection{Doppler-broadened susceptibility }
In the case of Doppler-broadened susceptibility,
the absorption profile near the center
(corresponding to zero velocity) is quite flat.
As the two windows occur near the center;
the maxima for both windows are the same.
The maximum value is calculated for
$\Omega_\text{c}=\Omega_\text{s}=0$ and for $\delta_\text{p}=\delta_\text{c}$:
\begin{equation}
	h_{\text{Dmax}_{1,2}}=\frac{\eta\sqrt{\pi}\left(\rho_{11}-\rho_{44}\right)}{\gamma_4+W_\text{L}}.
\end{equation}
The minimum value of the first window is calculated for
$\Omega_\text{s}=0$ and $\delta_\text{p}=\delta_\text{c}$:
\begin{equation}
	h_{\text{Dmin}_1}=\frac{\eta\sqrt{\pi}\left(\rho_{11}-\rho_{44}\right)\gamma_2}{\left|\Omega_\text{c}\right|^2+\gamma_2\left(\gamma_4+W_\text{L}\right)}
			\xrightarrow{\scriptscriptstyle \gamma_2\to 0} 0,
\end{equation}
which requires the condition $\left|\Omega_\text{c}\right|^2\gg\gamma_2\left(\gamma_4+W_\text{L}\right)$ 
to hold in order to reach minimum absorption.

For $\gamma_2\neq0$, however, for $W_\text{L}\gg\gamma_4$
this condition can be reduced to $\left|\Omega_\text{c}\right|^2\gg\gamma_2W_\text{L}$.
If the intensity of the driving field eliminates the inhomogeneous broadening due to Doppler broadening,
it certainly eliminates the homogeneous broadening as well.
The half-maximum of the first window is then equal to
\begin{equation}
	\bar{\varkappa}_1
		=\frac{\eta\sqrt{\pi}\left(\rho_{11}-\rho_{44}\right)}{2}\frac{2\gamma_2W_\text{L}+\left|\Omega_\text{c}\right|^2}{\gamma_2W^2_\text{L}+\left|\Omega_\text{c}\right|^2\left(\gamma_4+W_\text{L}\right)}.
\end{equation}

For $\imath=2$,
the minimum is calculated for $\Omega_\text{c}=0$ and $\delta_\text{p}=\delta_s$
with the result
\begin{align}
\label{minD21}
	h_{\text{Dmin}_2}
		=&\eta\sqrt{\pi}\left[\frac{\left(\rho_{11}-\rho_{44}\right)\gamma_3}{\gamma_3\left(\gamma_4+W_\text{L}\right)+\left|\Omega_\text{s}\right|^2}\right.\nonumber\\
	&-\frac{\left(\rho_{44}-\rho_{33}\right)\left|\Omega_\text{s}\right|^2}
		{\gamma_3 W^2_\text{L}+\left(W_\text{L}-\gamma_4\right)\left|\Omega_\text{s}\right|^2}
\nonumber\\
&\left.+\frac{2\left(\rho_{44}-\rho_{33}\right)\left|\Omega_\text{s}\right|^2}{ W_\text{L}\left(2\gamma_3\gamma_4+\left|\Omega_\text{s}\right|^2\right)}\right].
\end{align} 
The first term in the right-hand side of Eq.~(\ref{minD21}) represents the absorption minimum.
This term tends to~$0$ if $\gamma_3\rightarrow 0$.

For the case of nonzero dephasing or relaxation decay from state~$\ket{3}$,
condition $\left|\Omega_\text{s}\right|^2\gg\gamma_3\left(\gamma_4+W_\text{L}\right)$
is required to minimize the absorption.
For $W_\text{L}\gg\gamma_4$, this condition can be reduced to~(\ref{eq:inhom2}).
The last two terms of the right-hand side of Eq.~(\ref{minD21}) represent the maximum of the gain.
After solving some algebraic expressions,
Eq.~(\ref{minD21}) becomes 
\begin{align}
\label{minD22}
	h_{\text{Dmin}_2}
		=&\eta\sqrt{\pi}\left[\frac{\left(\rho_{11}-\rho_{44}\right)\gamma_3}{\gamma_3\left(\gamma_4+W_\text{L}\right)+\left|\Omega_\text{s}\right|^2}\right.\\ \nonumber
		&-\frac{\left(\rho_{44}-\rho_{33}\right)\left|\Omega_\text{s}\right|^2}{W_\text{L}(2\gamma_3\gamma_4+\left|\Omega_\text{s}\right|^2)}\\ \nonumber
&\left.\times\frac{2\gamma_3W_\text{L}(\gamma_4-W_\text{L})+\left|\Omega_\text{s}\right|^2(2\gamma_4-W_\text{L})}{\gamma_3W^2_\text{L}-\left|\Omega_\text{s}\right|^2(\gamma_4-W_\text{L})}\right]
\end{align}
Now, we want to examine whether condition~(\ref{eq:inhom2}),
for $\gamma_3\neq0$,
is required to observe gain of the Doppler broadening susceptibility.
If not,
then the existence of gain suppresses absorption,
and the second transparency window is observed even if condition~(\ref{eq:inhom2}) fails.
We evaluate Eq.~(\ref{minD22}) for the condition $W_\text{L}\gg\gamma_4$, 
in order to simplify the calculation,
and evaluate for condition~(\ref{eq:hom2}),
which is necessary to minimize the absorption as shown earlier:
\begin{align}
\label{minD24}
	h_{\text{Dmin}_2}
		=&\eta\sqrt{\pi}\left[\frac{\left(\rho_{11}-\rho_{44}\right)\gamma_3}{\gamma_3W_\text{L}+\left|\Omega_\text{s}\right|^2}\right.\\ \nonumber
	&\left.-\frac{\left(\rho_{44}-\rho_{33}\right)}{W_\text{L}}\left(1+\frac{\gamma_3W_\text{L}}{\gamma_3W_\text{L}+\left|\Omega_\text{s}\right|^2}\right)\right].
\end{align}
In order for gain to exist, 
\begin{align}
	\frac{\left|\rho_{44}-\rho_{33}\right|}{W_\text{L}}\left(1+\frac{\gamma_3W_\text{L}}{\gamma_3W_\text{L}+\left|\Omega_\text{s}\right|^2}\right)\gg\frac{\left|\rho_{11}-\rho_{44}\right|\gamma_3}{\gamma_3W_\text{L}+\left|\Omega_\text{s}\right|^2},
\end{align}
which can be simplified by rearranging terms
and substituting the quantity
\begin{equation}
	\frac{\left|\rho_{44}-\rho_{33}\right|}{\left|\rho_{11}-\rho_{44}\right|}\approx1
\end{equation}
to yield
\begin{equation}
	\frac{\left|\Omega_\text{s}\right|^2+W_\text{L}}{\gamma_3W_\text{L}}\gg0.
\label{ineq}
\end{equation}
Condition~(\ref{ineq}) is always valid even if condition~(\ref{eq:inhom2}) is not satisfied.
Note that the derivation of inequality~(\ref{ineq})
is based on the validity of condition~(\ref{eq:hom2}) for homogeneous broadening.
Therefore, condition~(\ref{eq:hom2}) is required for the gain to exist in our system, 
whereas condition~(\ref{eq:inhom2}) is not.

The half-maximum for the second EIT window,
without making any approximation,
is
\begin{align}
\label{HMD2}
	\bar{\varkappa}_2
		=&\frac{\eta\sqrt{\pi}}{2}
			\left[\frac{\left(\rho_{11}-\rho_{44}\right)\left(2\gamma_3\left(\gamma_4+W_\text{L}\right)+\left|\Omega_\text{s}\right|^2\right)}{\left(\gamma_3\left(\gamma_4+W_\text{L}\right)+\left|\Omega_\text{s}\right|^2\right)\left(\gamma_4+W_\text{L}\right)}\right.\nonumber\\
	&+\frac{\left(\rho_{44}-\rho_{33}\right)\left|\Omega_\text{s}\right|^2}{W_\text{L}(2\gamma_3\gamma_4+\left|\Omega_\text{s}\right|^2)}\nonumber\\
	&\left.\times\frac{2\gamma_3W_\text{L}(\gamma_4-W_\text{L})+\left|\Omega_\text{s}\right|^2(2\gamma_4-W_\text{L})}{-\gamma_3W^2_\text{L}+\left|\Omega_\text{s}\right|^2(\gamma_4-W_\text{L})}\right].
\end{align}
Applying condition~(\ref{eq:hom2}) simplifies this expression to
\begin{align}
\label{HMD22}
	\bar{\varkappa}_2
		=&\frac{\eta\sqrt{\pi}}{2}\left[\frac{\left(\rho_{11}-\rho_{44}\right)\left(2\gamma_3W_\text{L}+\left|\Omega_\text{s}\right|^2\right)}{\left(\gamma_3W_\text{L}+\left|\Omega_\text{s}\right|^2\right)\left(\gamma_4+W_\text{L}\right)}\right.\nonumber\\
	&+\left(\frac{2\gamma_3W_\text{L}(\gamma_4-W_\text{L})
		+\left|\Omega_\text{s}\right|^2(2\gamma_4-W_\text{L})}{-\gamma_3W^2_\text{L}+\left|\Omega_\text{s}\right|^2(\gamma_4-W_\text{L})}\right)\nonumber\\ 
	&\left.\times\frac{\rho_{44}-\rho_{33}}{W_\text{L}}\right],
\end{align}
which is the same as Eq.~(\ref{vaarkappa2}) for $\rho_{11}=\rho_{33}\approx0.5$ and $\rho_{44}=0$.

For the condition that $W_\text{L}\gg\gamma_4$,
Eq.~(\ref{HMD22}) reduces to
\begin{align}
\label{HMD23}
	\bar{\varkappa}_2
		=&\frac{\eta\sqrt{\pi}}{2W_\text{L}}\left(\frac{2\gamma_3W_\text{L}
			+\left|\Omega_\text{s}\right|^2}{\gamma_3W_\text{L}+\left|\Omega_\text{s}\right|^2}\right)\left(\rho_{11}-\rho_{33}\right).
\end{align}
Thus,
the half-maximum of the second EIT window for high Doppler broadening depends on the population difference between states~$\ket{1}$ and~$\ket{3}$.
For equal population,
the half maximum is always located at zero where absorption vanishes.
\label{sec:maximaminima}
\bibliography{Loren}

\begin{thebibliography}{22}
\expandafter\ifx\csname natexlab\endcsname\relax\def\natexlab#1{#1}\fi
\expandafter\ifx\csname bibnamefont\endcsname\relax
  \def\bibnamefont#1{#1}\fi
\expandafter\ifx\csname bibfnamefont\endcsname\relax
  \def\bibfnamefont#1{#1}\fi
\expandafter\ifx\csname citenamefont\endcsname\relax
  \def\citenamefont#1{#1}\fi
\expandafter\ifx\csname url\endcsname\relax
  \def\url#1{\texttt{#1}}\fi
\expandafter\ifx\csname urlprefix\endcsname\relax\def\urlprefix{URL }\fi
\providecommand{\bibinfo}[2]{#2}
\providecommand{\eprint}[2][]{\url{#2}}

\bibitem[{\citenamefont{Harris}(1997)}]{Harris1997}
\bibinfo{author}{\bibfnamefont{S.~E.} \bibnamefont{Harris}},
  \bibinfo{journal}{Phys. Today} \textbf{\bibinfo{volume}{50}},
  \bibinfo{pages}{(7), 36} (\bibinfo{year}{1997}).

\bibitem[{\citenamefont{Hau et~al.}(1999)\citenamefont{Hau, Harris, Dutton, and
  Behroozi}}]{Hau1999}
\bibinfo{author}{\bibfnamefont{L.~V.} \bibnamefont{Hau}},
  \bibinfo{author}{\bibfnamefont{S.~E.} \bibnamefont{Harris}},
  \bibinfo{author}{\bibfnamefont{Z.}~\bibnamefont{Dutton}}, \bibnamefont{and}
  \bibinfo{author}{\bibfnamefont{C.~H.} \bibnamefont{Behroozi}},
  \bibinfo{journal}{Nature (Londen)} \textbf{\bibinfo{volume}{397}},
  \bibinfo{pages}{594} (\bibinfo{year}{1999}).

\bibitem[{\citenamefont{Harris and Hau}(1999)}]{Harris99}
\bibinfo{author}{\bibfnamefont{S.~E.} \bibnamefont{Harris}} \bibnamefont{and}
  \bibinfo{author}{\bibfnamefont{L.~V.} \bibnamefont{Hau}},
  \bibinfo{journal}{Phys. Rev. Lett.} \textbf{\bibinfo{volume}{82}},
  \bibinfo{pages}{4611} (\bibinfo{year}{1999}).

\bibitem[{\citenamefont{Kash et~al.}(1999)\citenamefont{Kash, Sautenkov,
  Zibrov, Hollberg, Welch, Lukin, Rostovtsev, Fry, and Scully}}]{Kash1999}
\bibinfo{author}{\bibfnamefont{M.~M.} \bibnamefont{Kash}},
  \bibinfo{author}{\bibfnamefont{V.~A.} \bibnamefont{Sautenkov}},
  \bibinfo{author}{\bibfnamefont{A.~S.} \bibnamefont{Zibrov}},
  \bibinfo{author}{\bibfnamefont{L.}~\bibnamefont{Hollberg}},
  \bibinfo{author}{\bibfnamefont{G.~R.} \bibnamefont{Welch}},
  \bibinfo{author}{\bibfnamefont{M.~D.} \bibnamefont{Lukin}},
  \bibinfo{author}{\bibfnamefont{Y.}~\bibnamefont{Rostovtsev}},
  \bibinfo{author}{\bibfnamefont{E.~S.} \bibnamefont{Fry}}, \bibnamefont{and}
  \bibinfo{author}{\bibfnamefont{M.~O.} \bibnamefont{Scully}},
  \bibinfo{journal}{Phys. Rev. Lett.} \textbf{\bibinfo{volume}{82}},
  \bibinfo{pages}{5229} (\bibinfo{year}{1999}).

\bibitem[{\citenamefont{Harris and Yamamoto}(1998)}]{Harris1998}
\bibinfo{author}{\bibfnamefont{S.~E.} \bibnamefont{Harris}} \bibnamefont{and}
  \bibinfo{author}{\bibfnamefont{Y.}~\bibnamefont{Yamamoto}},
  \bibinfo{journal}{Phys. Rev. Lett.} \textbf{\bibinfo{volume}{81}},
  \bibinfo{pages}{3611} (\bibinfo{year}{1998}).

\bibitem[{\citenamefont{Lvovsky et~al.}(2009)\citenamefont{Lvovsky, Sanders,
  and Tittel}}]{Lvovsky2009}
\bibinfo{author}{\bibfnamefont{A.~I.} \bibnamefont{Lvovsky}},
  \bibinfo{author}{\bibfnamefont{B.~C.} \bibnamefont{Sanders}},
  \bibnamefont{and} \bibinfo{author}{\bibfnamefont{W.}~\bibnamefont{Tittel}},
  \bibinfo{journal}{Nat. Photon} \textbf{\bibinfo{volume}{3}},
  \bibinfo{pages}{706} (\bibinfo{year}{2009}).

\bibitem[{\citenamefont{Rebic et~al.}(2004)\citenamefont{Rebic, Vitali,
  Ottaviani, Tombesi, Artoni, Cataliotti, and Corbalan}}]{Rebic2004}
\bibinfo{author}{\bibfnamefont{S.}~\bibnamefont{Rebic}},
  \bibinfo{author}{\bibfnamefont{D.}~\bibnamefont{Vitali}},
  \bibinfo{author}{\bibfnamefont{C.}~\bibnamefont{Ottaviani}},
  \bibinfo{author}{\bibfnamefont{P.}~\bibnamefont{Tombesi}},
  \bibinfo{author}{\bibfnamefont{M.}~\bibnamefont{Artoni}},
  \bibinfo{author}{\bibfnamefont{F.}~\bibnamefont{Cataliotti}},
  \bibnamefont{and} \bibinfo{author}{\bibfnamefont{R.}~\bibnamefont{Corbalan}},
  \bibinfo{journal}{Phys. Rev. A} \textbf{\bibinfo{volume}{70}},
  \bibinfo{pages}{032 317} (\bibinfo{year}{2004}).

\bibitem[{\citenamefont{Alotaibi and Sanders}(2014)}]{Hessa2013}
\bibinfo{author}{\bibfnamefont{H.~M.~M.} \bibnamefont{Alotaibi}}
  \bibnamefont{and} \bibinfo{author}{\bibfnamefont{B.~C.}
  \bibnamefont{Sanders}}, \bibinfo{journal}{Phys. Rev. A}
  \textbf{\bibinfo{volume}{89}}, \bibinfo{pages}{021802}
  (\bibinfo{year}{2014}).

\bibitem[{\citenamefont{Schmidt and Imamoglu}(1996)}]{Schmidt1996}
\bibinfo{author}{\bibfnamefont{H.}~\bibnamefont{Schmidt}} \bibnamefont{and}
  \bibinfo{author}{\bibfnamefont{A.}~\bibnamefont{Imamoglu}},
  \bibinfo{journal}{Opt. Lett.} \textbf{\bibinfo{volume}{21 (23)}},
  \bibinfo{pages}{1936} (\bibinfo{year}{1996}).

\bibitem[{\citenamefont{Taichenachev et~al.}(2000)\citenamefont{Taichenachev,
  Tumaikin, and Yudin}}]{Taichenachev2000}
\bibinfo{author}{\bibfnamefont{A.~V.} \bibnamefont{Taichenachev}},
  \bibinfo{author}{\bibfnamefont{A.~M.} \bibnamefont{Tumaikin}},
  \bibnamefont{and} \bibinfo{author}{\bibfnamefont{V.~I.} \bibnamefont{Yudin}},
  \bibinfo{journal}{JETP Lett.} \textbf{\bibinfo{volume}{72}},
  \bibinfo{pages}{119} (\bibinfo{year}{2000}).

\bibitem[{\citenamefont{Rostovtsev et~al.}(2002)\citenamefont{Rostovtsev,
  Protsenko, Lee, and Javan}}]{Yuri2002}
\bibinfo{author}{\bibfnamefont{Y.}~\bibnamefont{Rostovtsev}},
  \bibinfo{author}{\bibfnamefont{I.}~\bibnamefont{Protsenko}},
  \bibinfo{author}{\bibfnamefont{H.}~\bibnamefont{Lee}}, \bibnamefont{and}
  \bibinfo{author}{\bibfnamefont{A.}~\bibnamefont{Javan}}, \bibinfo{journal}{J.
  Mod. Opt.} \textbf{\bibinfo{volume}{49}}, \bibinfo{pages}{2501}
  (\bibinfo{year}{2002}).

\bibitem[{\citenamefont{Javan et~al.}(2002)\citenamefont{Javan, Kocharovskaya,
  Lee, and Scully}}]{Javan2002}
\bibinfo{author}{\bibfnamefont{A.}~\bibnamefont{Javan}},
  \bibinfo{author}{\bibfnamefont{O.}~\bibnamefont{Kocharovskaya}},
  \bibinfo{author}{\bibfnamefont{H.}~\bibnamefont{Lee}}, \bibnamefont{and}
  \bibinfo{author}{\bibfnamefont{M.~O.} \bibnamefont{Scully}},
  \bibinfo{journal}{Phys. Rev. A} \textbf{\bibinfo{volume}{66}},
  \bibinfo{pages}{013805} (\bibinfo{year}{2002}).

\bibitem[{\citenamefont{Ye and Zibrov}(2002)}]{Ye2002}
\bibinfo{author}{\bibfnamefont{C.~Y.} \bibnamefont{Ye}} \bibnamefont{and}
  \bibinfo{author}{\bibfnamefont{A.~S.} \bibnamefont{Zibrov}},
  \bibinfo{journal}{Phys. Rev. A} \textbf{\bibinfo{volume}{65}},
  \bibinfo{pages}{023806} (\bibinfo{year}{2002}).

\bibitem[{\citenamefont{Figueroa et~al.}(2006)\citenamefont{Figueroa, Vewinger,
  Appel, and Lvovsky}}]{Figueroa2006}
\bibinfo{author}{\bibfnamefont{E.}~\bibnamefont{Figueroa}},
  \bibinfo{author}{\bibfnamefont{F.}~\bibnamefont{Vewinger}},
  \bibinfo{author}{\bibfnamefont{J.}~\bibnamefont{Appel}}, \bibnamefont{and}
  \bibinfo{author}{\bibfnamefont{A.~I.} \bibnamefont{Lvovsky}},
  \bibinfo{journal}{Opt. Lett.} \textbf{\bibinfo{volume}{31}},
  \bibinfo{pages}{2625} (\bibinfo{year}{2006}).

\bibitem[{\citenamefont{Fleischhauer et~al.}(2005)\citenamefont{Fleischhauer,
  Imamoglu, and Marangos}}]{Fleischhauer2005}
\bibinfo{author}{\bibfnamefont{M.}~\bibnamefont{Fleischhauer}},
  \bibinfo{author}{\bibfnamefont{A.}~\bibnamefont{Imamoglu}}, \bibnamefont{and}
  \bibinfo{author}{\bibfnamefont{J.~P.} \bibnamefont{Marangos}},
  \bibinfo{journal}{Rev. Mod. Phys.} \textbf{\bibinfo{volume}{77}},
  \bibinfo{pages}{633} (\bibinfo{year}{2005}).

\bibitem[{\citenamefont{Gea-Banacloche
  et~al.}(1995)\citenamefont{Gea-Banacloche, Li, Jin, and Xiao}}]{Julio1995}
\bibinfo{author}{\bibfnamefont{J.}~\bibnamefont{Gea-Banacloche}},
  \bibinfo{author}{\bibfnamefont{Y.-q.} \bibnamefont{Li}},
  \bibinfo{author}{\bibfnamefont{S.-z.} \bibnamefont{Jin}}, \bibnamefont{and}
  \bibinfo{author}{\bibfnamefont{M.}~\bibnamefont{Xiao}},
  \bibinfo{journal}{Phys. Rev. A} \textbf{\bibinfo{volume}{51}},
  \bibinfo{pages}{576} (\bibinfo{year}{1995}).

\bibitem[{\citenamefont{Pagnini and Saxena}(2008)}]{Pag08}
\bibinfo{author}{\bibfnamefont{G.}~\bibnamefont{Pagnini}} \bibnamefont{and}
  \bibinfo{author}{\bibfnamefont{R.~K.} \bibnamefont{Saxena}}
  (\bibinfo{year}{2008}), \bibinfo{note}{arXiv:math-phys}, \eprint{0805.2274}.

\bibitem[{\citenamefont{Vemuri and Agarwal}(1996)}]{VA96}
\bibinfo{author}{\bibfnamefont{G.}~\bibnamefont{Vemuri}} \bibnamefont{and}
  \bibinfo{author}{\bibfnamefont{G.~S.} \bibnamefont{Agarwal}},
  \bibinfo{journal}{Phys. Rev. A} \textbf{\bibinfo{volume}{53}},
  \bibinfo{pages}{1060} (\bibinfo{year}{1996}).

\bibitem[{\citenamefont{Li et~al.}(2007)\citenamefont{Li, Yang, Cao, Zhang,
  Xie, and Wang}}]{Li2007}
\bibinfo{author}{\bibfnamefont{S.}~\bibnamefont{Li}},
  \bibinfo{author}{\bibfnamefont{X.}~\bibnamefont{Yang}},
  \bibinfo{author}{\bibfnamefont{X.}~\bibnamefont{Cao}},
  \bibinfo{author}{\bibfnamefont{C.}~\bibnamefont{Zhang}},
  \bibinfo{author}{\bibfnamefont{C.}~\bibnamefont{Xie}}, \bibnamefont{and}
  \bibinfo{author}{\bibfnamefont{H.}~\bibnamefont{Wang}}, \bibinfo{journal}{J.
  Phys. B: At. Mol. Opt. Phys.} \textbf{\bibinfo{volume}{40}},
  \bibinfo{pages}{3211} (\bibinfo{year}{2007}).

\bibitem[{\citenamefont{MacRae et~al.}(2008)\citenamefont{MacRae, Campbell, and
  Lvovsky}}]{MCL08}
\bibinfo{author}{\bibfnamefont{A.}~\bibnamefont{MacRae}},
  \bibinfo{author}{\bibfnamefont{G.}~\bibnamefont{Campbell}}, \bibnamefont{and}
  \bibinfo{author}{\bibfnamefont{A.~I.} \bibnamefont{Lvovsky}},
  \bibinfo{journal}{Opt. Lett} \textbf{\bibinfo{volume}{33}},
  \bibinfo{pages}{2659} (\bibinfo{year}{2008}).

\bibitem[{\citenamefont{Ackerhalt et~al.}(1979)\citenamefont{Ackerhalt, Eberly,
  and Shore}}]{Ackerhalt79}
\bibinfo{author}{\bibfnamefont{J.~R.} \bibnamefont{Ackerhalt}},
  \bibinfo{author}{\bibfnamefont{J.~H.} \bibnamefont{Eberly}},
  \bibnamefont{and} \bibinfo{author}{\bibfnamefont{B.~W.} \bibnamefont{Shore}},
  \bibinfo{journal}{Phys. Rev. A} \textbf{\bibinfo{volume}{19}},
  \bibinfo{pages}{248} (\bibinfo{year}{1979}).

\bibitem[{\citenamefont{Choe et~al.}(1995)\citenamefont{Choe, Han, Rhee, Lee,
  Han, Kuzmina, and Mishin}}]{Choe1995}
\bibinfo{author}{\bibfnamefont{A.~S.} \bibnamefont{Choe}},
  \bibinfo{author}{\bibfnamefont{J.}~\bibnamefont{Han}},
  \bibinfo{author}{\bibfnamefont{Y.}~\bibnamefont{Rhee}},
  \bibinfo{author}{\bibfnamefont{J.}~\bibnamefont{Lee}},
  \bibinfo{author}{\bibfnamefont{P.~S.} \bibnamefont{Han}},
  \bibinfo{author}{\bibfnamefont{M.~A.} \bibnamefont{Kuzmina}},
  \bibnamefont{and} \bibinfo{author}{\bibfnamefont{V.~A.}
  \bibnamefont{Mishin}}, \bibinfo{journal}{J. Phys. B: At. Mol. Opt. Phys}
  \textbf{\bibinfo{volume}{28}}, \bibinfo{pages}{2355} (\bibinfo{year}{1995}).

\end{thebibliography}
\end{document}